\documentclass[apj,numberedappendix]{emulateapj}	
\pdfoutput=1

\usepackage[
debug=true,
hyperindex=true,
colorlinks=false,
bookmarksnumbered=true,
baseurl={http://www.cfa.harvard.edu},
hyperfootnotes=false, 
pdfkeywords={Dust, Extinction, Pipe Nebula, Clouds, Dynamics, Star formation},
pdftitle={The nature of the dense core population in the pipe nebula:
core and cloud kinematics from C18O observations},
pdfauthor={August Muench}]{hyperref} 
\journalinfo{To appear in the Astrophysical Journal}%
\submitted{Submitted  11 Jun 2007; Accepted  04 Sep 2007; VERSION \today}%

\slugcomment{VERSION \today; Submitted June 11, 2007}
\shorttitle{\cotracer\ observations of the Pipe nebula}
\shortauthors{Muench et al.}

\newcommand{\mass}{\ensuremath{ \mathcal{M} }}
\newcommand{\solarmass}{\ensuremath{ \mathcal{M}_{\Sun} }}
\newcommand{\av}{\ensuremath{\mathnormal{A}_{V}}}
\newcommand{\cotracer}{\ensuremath{\mbox{C}^{18}\mbox{O}}}
\newcommand{\comap}{\ensuremath{{\ ^{13}}{\mbox{CO}}}}
\newcommand{\vlsr}{\ensuremath{{v}_{lsr}}}
\newcommand{\dvco}{\ensuremath{{{\Delta}v}}}
\newcommand{\tant}{\ensuremath{{T}_{R}^{*}}}
\newcommand{\kms}{km~s$^{-1}$}

\begin{document}
 
\title{The nature of the dense core population in the pipe nebula:\\
core and cloud kinematics from \cotracer\ observations
}%

\author{August A. Muench\altaffilmark{1},
		Charles J. Lada\altaffilmark{1},
		Jill M. Rathborne\altaffilmark{1}, \\
		Jo\~{a}o F. Alves\altaffilmark{2},
		\&
		M. Lombardi\altaffilmark{3,4}}

\altaffiltext{1}{Smithsonian Astrophysical Observatory.
60 Garden Street, Mail Stop 72, Cambridge, MA~~02138~USA;
~gmuench@cfa.harvard.edu}
\altaffiltext{2}{Calar Alto Observatory. 
Centro Astron\'{o}mico Hispano Alem\'{a}n, 
c/Jes\'{u}s Durb\'{a}n Rem\'{o}n 2-2, 04004 Almeria, Spain
}
\altaffiltext{3}{European Southern Observatory. 
Karl-Schwarzschild-Str. 2, 85748 Garching, Germany
}
\altaffiltext{4}{University of Milan, Department of Physics. 
via~Celoria 16, 20133 Milan, Italy
}


%
\begin{abstract}
We present molecular-line observations of 94 dark cloud cores identified in
the Pipe nebula through near-IR extinction mapping. Using the Arizona Radio
Observatory 12m telescope, we obtained spectra of these
cores in the J=1-0 transition of \cotracer. We used the measured core
parameters, \tant, \dvco, \vlsr, radius and mass, to explore the internal
kinematics of the cores as well as their radial motions through the larger
molecular cloud. We find that the vast majority of the dark extinction cores 
are true cloud cores rather than the superposition of unrelated filaments. 
While we identify no significant correlations between the core's internal gas motions 
and the cores' other physical parameters, we identify spatially correlated radial
velocity variations that outline two main kinematic components of the cloud. 
The largest is a 15~pc long filament that is surprisingly narrow both in 
spatial dimensions and in radial velocity. Beginning in the Stem of the Pipe, 
this filament displays uniformly small \cotracer\ linewidths 
$(\dvco\sim0.4\mbox{\kms})$ as well as core to core motions only slightly 
in excess of the gas sound speed. The second component outlines what appears 
to be part of a large $(2\mbox{pc};\;10^{3}\solarmass)$ ring-like structure.  
Cores associated with this component display both larger linewidths and 
core to core motions than in the main cloud. The Pipe Molecular Ring
may represent a primordial structure related to the formation of this cloud.  
\end{abstract}
\keywords{
 Dust, extinction ---
 ISM: individual (Pipe Nebula) ---
 ISM: clouds ---
 ISM: kinematics and dynamic ---
 radio lines: ISM ---
 stars: formation
}%

\section{Introduction}
\label{sec:intro}

The \object{Pipe nebula} is a nearby (130~pc), large (15~pc) dark molecular cloud 
lying in projection against the Galactic bulge $(l\sim0\degr; b\sim5\degr)$ and visible
to the unaided eye. In addition to being one of the most nearby dark clouds it is
unique for its apparent lack of star formation. Comparing large scale CO maps
to IRAS point sources, \citet{1999PASJ...51..871O} found only a single site of
star formation \citep[\object{Barnard~59};][]{1996A&A...314..258R, 2007ApJ...655..364B}
although their observations also revealed numerous dense (\cotracer) cores of gas.
Given this lack of star
formation, the Pipe nebula represents a good target for improving our
understanding of the formation and early evolution of dark clouds before the
star formation process has confused or significantly modified a cloud's
internal structure or kinematics. Indeed, the Pipe cloud probably lies at the
threshold before star formation begins and its internal structure and
kinematics represent the initial conditions of star formation needed to guide
theoretical models.

The physical structure of a dark cloud is best revealed through measurements
of the dust; in particular through the mapping of the extinguishing of the
starlight of background stars by this dust \citep{1923AN....219..109W,
1994ApJ...429..694L, 1999A&A...345..965C, 2001Natur.409..159A,
2005PASJ...57S...1D}. \citet{2006A&A...454..781L} used a multi-band near-IR
technique \citep[NICER;][]{2001A&A...377.1023L} to create a detailed
extinction map of the Pipe nebula at a resolution of $\sim1\arcmin$.
Filamentary structures with column densities as low as $\av\sim1$ mag 
can be seen weaving through the Pipe and this extinction map has a spatial
resolution and a density contrast ideal for investigating primordial cloud structure. 
\citet{2007A&A...462L..17A} combined a wavelet analysis of the cloud
with a clump finding routine to identify 159 significant column density
enhancements that we will term dark extinction ``cores.'' This terminology follows
the literature usage of identifying dark regions of optical extinction
\citep[e.g.][]{1999ApJS..123..233L} or dust continuum 
peaks \citep[e.g.][]{1998A&A...336..150M}
as the cores of molecular clouds.  We note that there is no consensus on
the defining characteristics of a cloud core; some researchers in the field
suggest that the terminology of a ``core'' should be applied only to
those objects that are gravitationally bound.  For this work we
retain the use of the term ``core,'' while also noting that such a 
determination cannot be made from dust absorption measurements alone.

Molecular-line observations are important to confirm that extinction
cores are dense coherent kinematic structures and not the random superposition
of cloud features such as filaments or multiple cloud cores.
Molecular-line observations trace both ``microscopic'' internal motions such
as turbulence, infall, rotation and expansion and any ``macroscopic'' bulk
mass motions of the dark cores through the large $(10^{4}\solarmass)$ cloud.
Thus, gas tracers are very useful for extracting the kinematics of dark clouds.
Here we use molecular gas tracers to explore the kinematic state of the Pipe
extinction cores identified in \citet{2007A&A...462L..17A}.
Specifically, we use observations of \cotracer\ at 3mm to measure the extinction
cores' non-thermal internal motions as well as their ballistic motion through the
ambient cloud traced by their radial velocity. This paper is the first in a
series exploring the nature of the Pipe extinction cores by integrating
tracers of molecular gas at a range of densities with the structure provided
by the extinction map. In \citet{JRathborne2007Pipe2} we use ammonia 
observations to explore the densest gas in the cores, deriving kinetic gas
temperatures and probing the cores' chemical state. Finally,
\citet{CLada2007Pipe3} provides a synthesis of these results and examines
their implications for the ability of these extinction cores to form young
stars and, thus, provide an origin for the stellar initial mass function. 
This paper is organized simply into three subsequent sections.
First, we detail our collection of \cotracer\ data for 94 of the 159 Pipe
extinction cores identified by \citeauthor{2007A&A...462L..17A}
(\S\ref{sec:data}), finding the vast majority to be coherent cloud structures.
In \S\ref{sec:cores} we analyze the completeness and
homogeneity of our dataset, and we explore correlations between the extinction
cores' structure and internal gas kinematics. Finally, we use our data to
examine the large scale motions of the cores through the Pipe cloud
(\S\ref{sec:discuss}), revealing large (10~pc) coherent structures, 
including a large Ring similar to those found in the Taurus star
forming region.

\section{Data}
\label{sec:data}

\subsection{Telescope}
\label{sec:obs}

Our goal was to survey the molecular gas properties of dark extinction cores
identified by \citet{2007A&A...462L..17A} in the 2MASS NICER map of
\citet{2006A&A...454..781L}. To complete this goal we used the Arizona Radio
Observatory Kitt Peak 12m telescope (hereafter the ARO12m) for 12 nights over
a 2 year period (2005-2006). We observed single center pointings of the J=1-0
line of \cotracer\ for each core. Table \ref{tab:obslog} lists the dates of
our observations as well as the typical system temperatures $(T_{sys})$ and
the number of pointings achieved per night; from Tucson, Arizona the Pipe
nebula has a maximum visibility (el $>$ 10\degr) window less than 5~hrs. All
observations were performed remotely from Cambridge, Massachusetts, USA.
%
\begin{deluxetable}{ccrccr}
\tablewidth{0pt}
\tablecaption{Observing Log\label{tab:obslog}}
\tablehead{
\multicolumn{3}{c}{2005} & 
\multicolumn{3}{c}{2006} \\
\colhead{Date-Obs} &
\colhead{$T_{sys}$} &
\colhead{$N_{obs}$} &
\colhead{Date-Obs} &
\colhead{$T_{sys}$} &
\colhead{$N_{obs}$}
}
\startdata
2005-05-10 & 270-370 & 19 & 2006-01-28 & 240-280 & 8  \\
2005-05-11 & 280-360 & 23 & 2006-01-29 & 310-430 & 9  \\
2005-05-12 & 240-330 & 36 & 2006-01-30 & 240-440 & 10 \\
2005-05-19 & 340-490 & 18 & 2006-01-31 & 210-290 & 7  \\
2005-06-25 & 490-700 & 11 & 2006-02-01 & 230-270 & 9  \\
           &         &    & 2006-02-02 & 240-290 & 10 
\enddata
\end{deluxetable}%

\subsection{Technique}
\label{sec:tel}

We next summarize the observational technique used to obtain the \cotracer\
data with the ARO12m. Observations were performed using the 3mm receiver
(90-116 GHz) operating in frequency switching mode with the millimeter auto
correlator. The correlator was configured in dual-polarization mode with 75
MHz usable bandwidth in each of 2 IF banks, having 16384 channels and an
effective spectral resolution of 24.4~KHz/channel or 0.067~\kms\ at 109~GHz.
We frequency switched at 5~Hz with a 2~MHz throw; the cloud appears projected against the
Galactic bulge and we did not expect nor did we observe any molecular gas not
associated with nearby gas clouds ($\vlsr$ between 0-20~\kms).
Spectral observations at the ARO12m use chopper-wheel calibration to correct
for atmospheric attenuation and telescope losses. This places
our data on the $\tant$ temperature scale \citep{1981ApJ...250..341K}.
The source main beam brightness
temperature, ${T}_{mb}$, is related to $\tant$ by $\tant/\eta_{m}^{*}$,
where $\eta_{m}^{*}=0.97$ at 109GHz at the ARO12m, 
assuming that the source fills the main beam.

The standard observation had an integration time of 10 minutes. For each target,
we repeated the standard observation until we achieved a root mean square
noise of $<0.1$ K per channel; total integrations ranged from 10 to 60 minutes
and this value depended more on the target's elevation than its \cotracer\
brightness. All Pipe extinction cores in our sample were detected (at $3\sigma$ or
greater) in \cotracer\ with the weakest line having $\tant=0.14K$~(Pipe-100).
Individual spectra were summed, folded, baseline subtracted and Gaussian
profile fit using CLASS. The reduced results assume a rest frame frequency for
\cotracer\ J=1-0 of 109782.173 MHz; this value was determined empirically by
\citet{1999ApJ...526..788L,2001ApJS..136..703L}.
%
\begin{figure}
\centering \includegraphics[angle=90,width=0.45\textwidth]{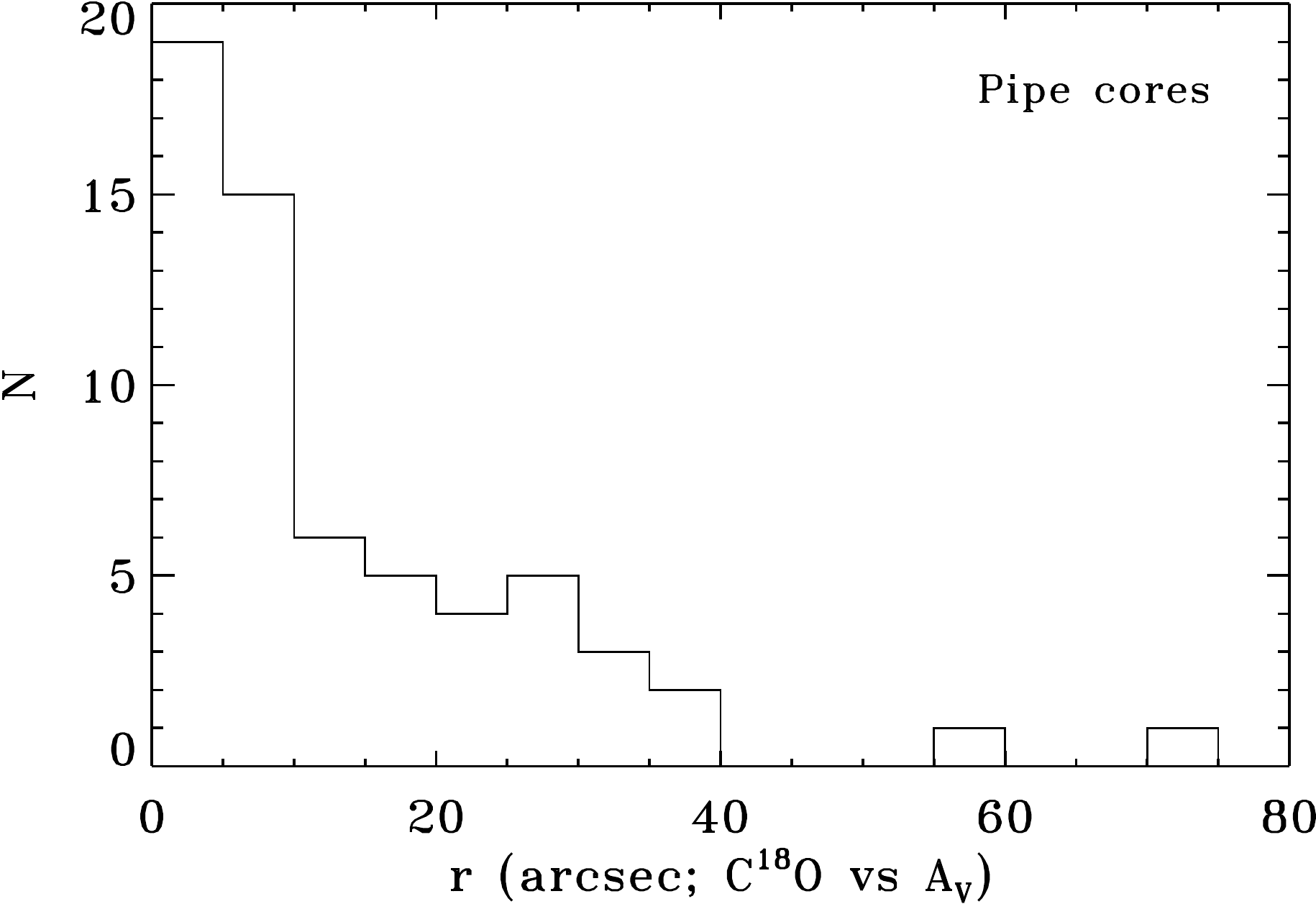}
\caption{
Positional offsets $(r_{sep})$ between the ARO12m \cotracer\ $56\arcsec$ beam and 
the extinction core position reported in \citet{2007A&A...462L..17A}.
Note, the ARO12m \cotracer\ beam was centered on the extinction peak
of each core.
\label{fig:offsets}}
\end{figure}%
%
\begin{deluxetable*}{rcllrrrrrr}
\tablecaption{\cotracer\ data\label{tab:c18o}}
\tablehead{
\colhead{Pipe} &
\colhead{flag} &
\multicolumn{2}{c}{Position (J2000)} &
\colhead{$r_{sep}$} &
\colhead{$v_{lsr}$} &
\colhead{$dv$} &
\colhead{$T_{R}^{*}$} &
\colhead{rms} & 
\colhead{$\Sigma\;\av$\tablenotemark{(b)}} \\
\colhead{(ID)} &
\colhead{\tablenotemark{(a)}} &
\colhead{RA} &
\colhead{DEC} &
\colhead{$(\arcsec)$} &
\multicolumn{2}{c}{(\kms)} &
\multicolumn{2}{c}{(K)} &
\colhead{(mag.)} 
}
\startdata
    6 &     & 17:10:31.01     & -27:25:33.31    &    8.07 &   3.501 &   0.391 &   2.066 &    0.06 &    99.6  \\
    7 &     & 17:11:36.39     & -27:33:50.70    &   19.97 &   3.935 &   0.517 &   1.366 &    0.05 &    85.6  \\
    8 &     & 17:12:15.04     & -27:37:44.41    &    2.76 &   3.462 &   0.390 &   2.012 &    0.05 &    88.7  \\
   11 &     & 17:10:49.87     & -27:23:04.07    &    6.06 &   3.446 &   0.455 &   1.524 &    0.07 &    85.2  \\
   13 &     & 17:10:47.22     & -27:13:41.43    &    9.64 &   3.751 &   0.336 &   0.498 &    0.05 &    39.4  \\
\enddata
\tablenotetext{(a)}{Flag given to indicate \cotracer\ component for each extinction core. Components ``a'', ``b'', etc
are ordered by antenna temperature. Where given, a Flag=2, corresponds to all components fit with a single
Gaussian.}
\tablenotetext{(b)}{Total column density expressed in magnitudes of extinction (\av) measured in a Gaussian
weighted 56\arcsec\ beam on the NICER 2MASS map of \citet{2006A&A...454..781L} and centered on the 
position of the \cotracer\ observation.}
\tablecomments{Full version given at end of document (Table \ref{tab:c18ofull}).}
\end{deluxetable*}%

\subsection{Targets}
\label{sec:targets}

In this section we describe the targets selected for these observations. The
56\arcsec\ beam of the ARO12m at 109 GHz is well matched to the Gaussian
kernel used by \citet{2006A&A...454..781L} to filter spatially individual
star extinction estimates derived from the 2MASS point source catalog. The
final 2MASS extinction map has a resolution of 60\arcsec\ FWHM Gaussian and
was Nyquist sampled to a pixel scale of 30\arcsec. Additionally, the ARO12m
beam is much smaller than the $2.7\arcmin$ beam used by
\citet{1999PASJ...51..871O} to create under-sampled \cotracer\ maps of
${}^{13}$CO cores (differences in beam size and resolution are discussed in
\S\ref{sec:dvsize}). Given this optimized beam we targeted localized peaks of
interstellar extinction throughout the Pipe nebula and assigned priority based
initially on maximum column density. We revised our target priorities after
the first year's observations to sample lower mass dark cores. In summary we
obtained single pointing \cotracer\ data for 94 of the 159 Pipe extinction
cores; these targets span the entire range of core mass \citep[0.2 to 20
\solarmass;][]{2007A&A...462L..17A}. The completeness and homogeneity of this
sample of cores are examined in \S\ref{sec:cores}.

While we targeted the localized extinction peak within each Pipe core, our
pointing centers correspond very closely to the core centers reported in
\citet{2007A&A...462L..17A}\footnote{The centers for the Pipe extinction
cores listed in Alves et al. were derived using a 2D version of Clumpfind
\citep{1994ApJ...428..693W}.}, 
lying typically within a single  resolution element of the $\av$ map. 
The distribution function of separations  between the published core 
centers and the \cotracer\ pointings is given in Figure~\ref{fig:offsets}.

\subsection{Results}
\label{sec:results}

We provide \cotracer\ observations for 94 Pipe extinction cores in Table
\ref{tab:c18o}. Identifications correspond to the core ID given by
\citet{2007A&A...462L..17A}\footnote{The \citeauthor{2007A&A...462L..17A}
extinction core identifications and positions can be obtained using the VizieR
service \url{http://vizier.cfa.harvard.edu/} and have the catalog id of
\protect{J/A+A/462/L17}.} and are identical to those given 
in \citet{CLada2007Pipe3}.  The Table includes the pointing center for each
\cotracer\ observation and the results of a Gaussian profile fit.

Nine extinction cores $(10\%\mbox{ of the sample})$ displayed multi-peaked profiles in
\cotracer; four of these cores consist of two and in one case three distinct
lines. We record in the Table individual Gaussian fits to each component,
identifying the stronger(\tant) line as the ``a'' cloud. In four of the
nine cases the lines are sufficiently blended that in addition to
multicomponent fits we also tabulated the fit of a single Gaussian to the
double-peaked profile. We further analyze kinematically distinct and
overlapping cloud components in \S\ref{sec:cloud}.

\section{Analysis}
\label{sec:cores}

\subsection{Completeness}
\label{sec:complete}

\begin{figure}
\centering \includegraphics[angle=90,width=0.45\textwidth]{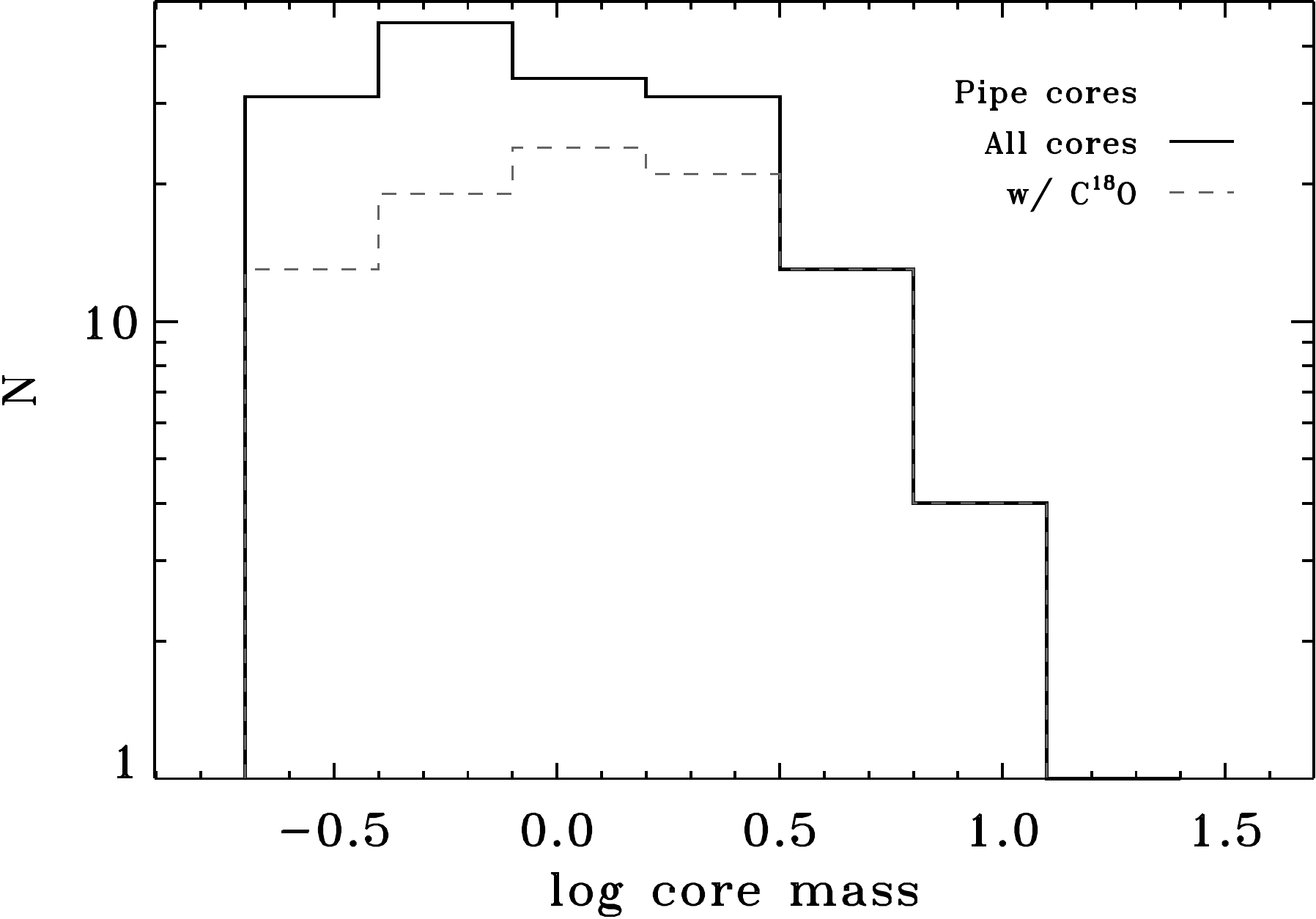}
\caption{
Completeness of CO sample. Mass functions for all 159 extinction cores 
(solid line) and for extinction cores with \cotracer\ measurements 
(dashed line) are compared.  Masses are taken from the wavelet subtracted 
map as tabulated in \citet[][]{CLada2007Pipe3}.
\label{fig:complete}}
\end{figure}

In this section we examine the completeness as a function of extinction 
core mass for our sample of 94 cores for which we obtained \cotracer\ line
measurements. In Figure~\ref{fig:complete} we compare the mass function of all
159 Pipe cores to the mass function for our sample with \cotracer\ data. Note
that the masses $(\mass_{core})$ and radii of the extinction cores used in 
this paper were derived by and are found in \citet{CLada2007Pipe3}. 
Briefly, Lada et al integrated the background subtracted extinction map 
of \citet{2007A&A...462L..17A} for cores with positions and sizes determined
by the 2D version of Clumpfind \citep{1994ApJ...428..693W}.  
We list in Table \ref{tab:complete} our completeness statistics. 
Our sample includes 100\% of the extinction cores with masses
greater than about $3\solarmass$ with the sole exception of the star
forming core Barnard~59 $(\mbox{Pipe-12}; \mass=20\solarmass)$. Between
$1\mbox{ and }3\solarmass$ our sample is $\sim70\%$ complete. Below 1
$\solarmass$ our sample includes $\sim40\%$ of all the
\citeauthor{2007A&A...462L..17A} Pipe extinction cores.
%
%
\begin{deluxetable}{rrrrrrr}
\tablewidth{0pt}
\tablecaption{\cotracer\ core completeness\label{tab:complete}}
\tablehead{
\colhead{$\log \solarmass$} &
\multicolumn{3}{c}{Full catalog} &
\multicolumn{3}{c}{Stem and Bowl} \\
\colhead{} &
\colhead{$N_{cores}$} &
\colhead{$N_{obs}$} &
\colhead{$\%$} &
\colhead{$N_{cores}$} &
\colhead{$N_{obs}$} &
\colhead{$\%$}
}
\startdata
-0.550 &  31 &  13 &     42. &  17 &  12 &     71. \\
-0.250 &  45 &  19 &     42. &  21 &  16 &     76. \\
 0.050 &  34 &  24 &     71. &  23 &  20 &     87. \\
 0.350 &  31 &  21 &     68. &  18 &  15 &     81. \\
 0.650 &  13 &  13 &    100. &  13 &  13 &    100. \\
 0.950 &   4 &   4 &    100. &   4 &   4 &    100. \\
 1.250 &   1 &   0 &      0. &   1 &   0 &      0. 
\enddata
\end{deluxetable}%
%
\begin{figure*}
\centering \includegraphics[angle=90,width=0.75\textwidth]{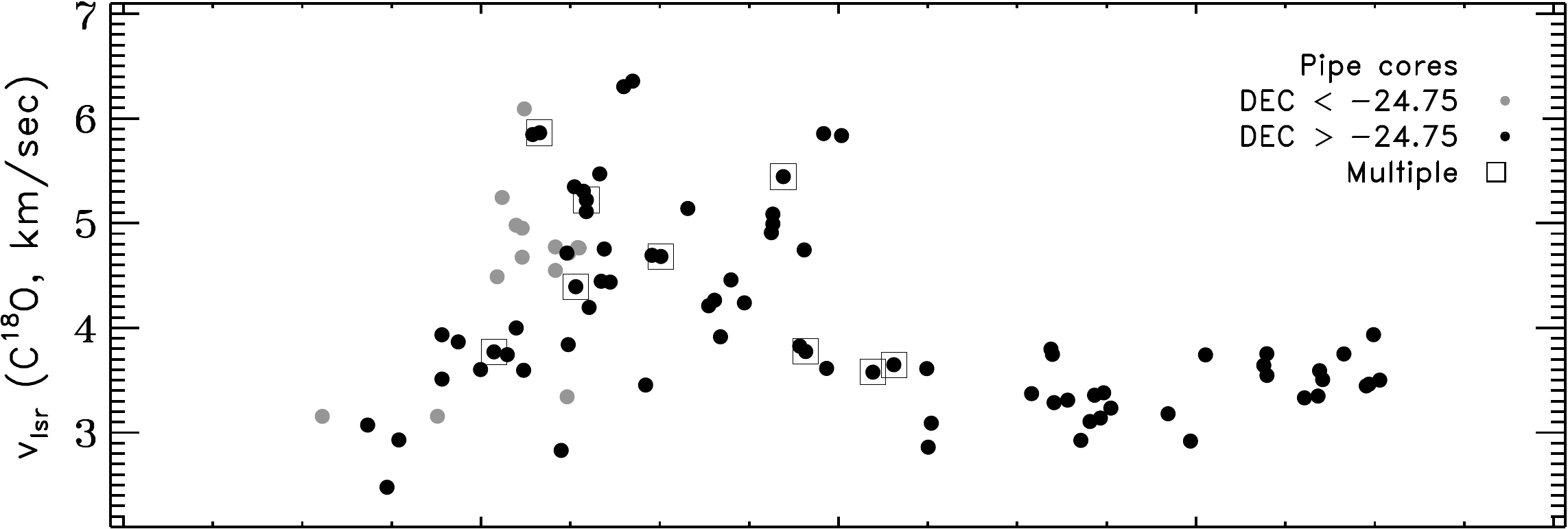}
\centering \includegraphics[angle=00,width=0.80\textwidth]{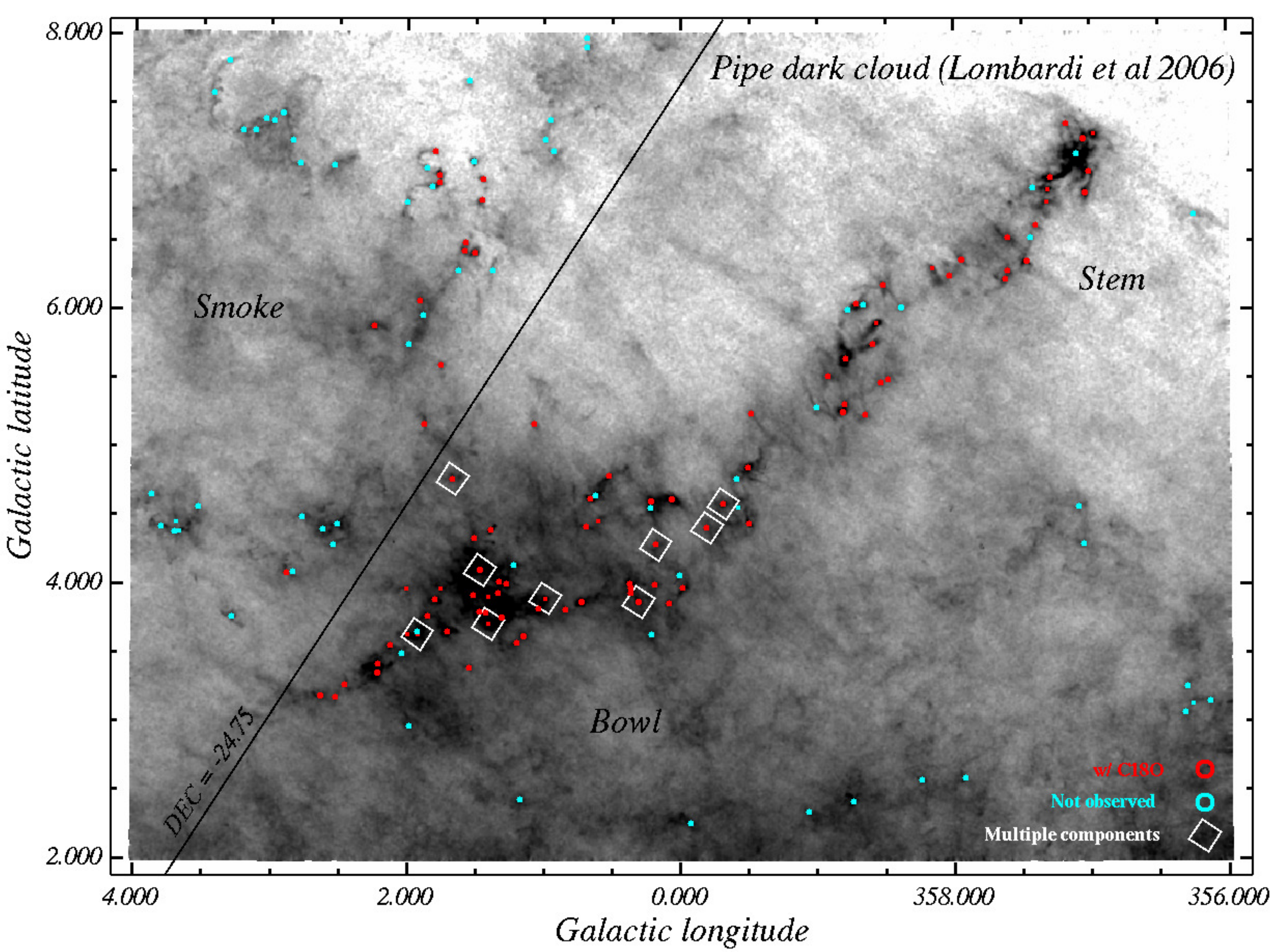}
\caption{
Spatial distribution of Pipe extinction cores.
$Top:$~Spatial variations in \vlsr\ of the Pipe cores observed
in \cotracer.
$Bottom:$~Spatial distribution of Pipe cores with and without 
\cotracer\ measurements. 
Underlying reverse grayscale image of the Pipe nebula in 
extinction \citep{2006A&A...454..781L}.
Three subregions of the Pipe referred to in text (Stem, Bowl, Smoke)
and used to characterize the sample are so labeled. 
Sources surrounded by diamonds display multiple \cotracer\ 
line components. Bottom~figure created using 
SAOImage DS9 \citep{2003ASPC..295..489J}. 
\label{fig:map1}}
\end{figure*}%

Our observations of the Pipe extinction cores are spatially non-uniform. We
plot in Figure~\ref{fig:map1}$b$ the 2MASS/NICER extinction map from
\citet{2006A&A...454..781L}, segregating cores with and without \cotracer\ by
color. For the main body of the Pipe $(DEC\,<\,-24.75\degr)$ we achieved a degree
of completeness much higher than in the composite core population: $>80\%$ for
1-3\solarmass cores and $>70\%$ of the subsolarmass cores. In the next section
we explore the spatial homogeneity of the properties of the observed cores.

\subsection{Spatial homogeneity}
\label{sec:represent}

In this section we evaluate our sample of \cotracer\ observed extinction
cores, comparing internally subsets of the data that are defined by their
location in the cloud. Spatial subsets provide a means to evaluate our
somewhat non-uniform spatial coverage; however, our spatial partitioning of
the Pipe nebula is further motivated by clear physical properties of the
cloud.

In panel Fig.~\ref{fig:map1}$a$ we illustrate the dependence of a core's
radial velocity with Galactic longitude. The range of observed \vlsr, 3 to
7~\kms, is consistent with that found for \cotracer\ cores throughout
Ophiuchus \citep{2000ApJ...528..817T} although we do not observe any core with
$\vlsr\ < 2$~\kms, for which \citeauthor{2000ApJ...528..817T} observed
several\footnote{Two cores having $\vlsr>7$~\kms, specifically at 9 and
15~\kms\ do not appear on Fig.~\ref{fig:map1} and are not further examined in
this work.}. But unlike Ophiuchus the spatial distribution of \vlsr\ is highly
structured, displaying strong correlations with Galactic longitude.

%
\begin{figure*}
\centering \includegraphics[angle=00,width=0.63\textwidth]{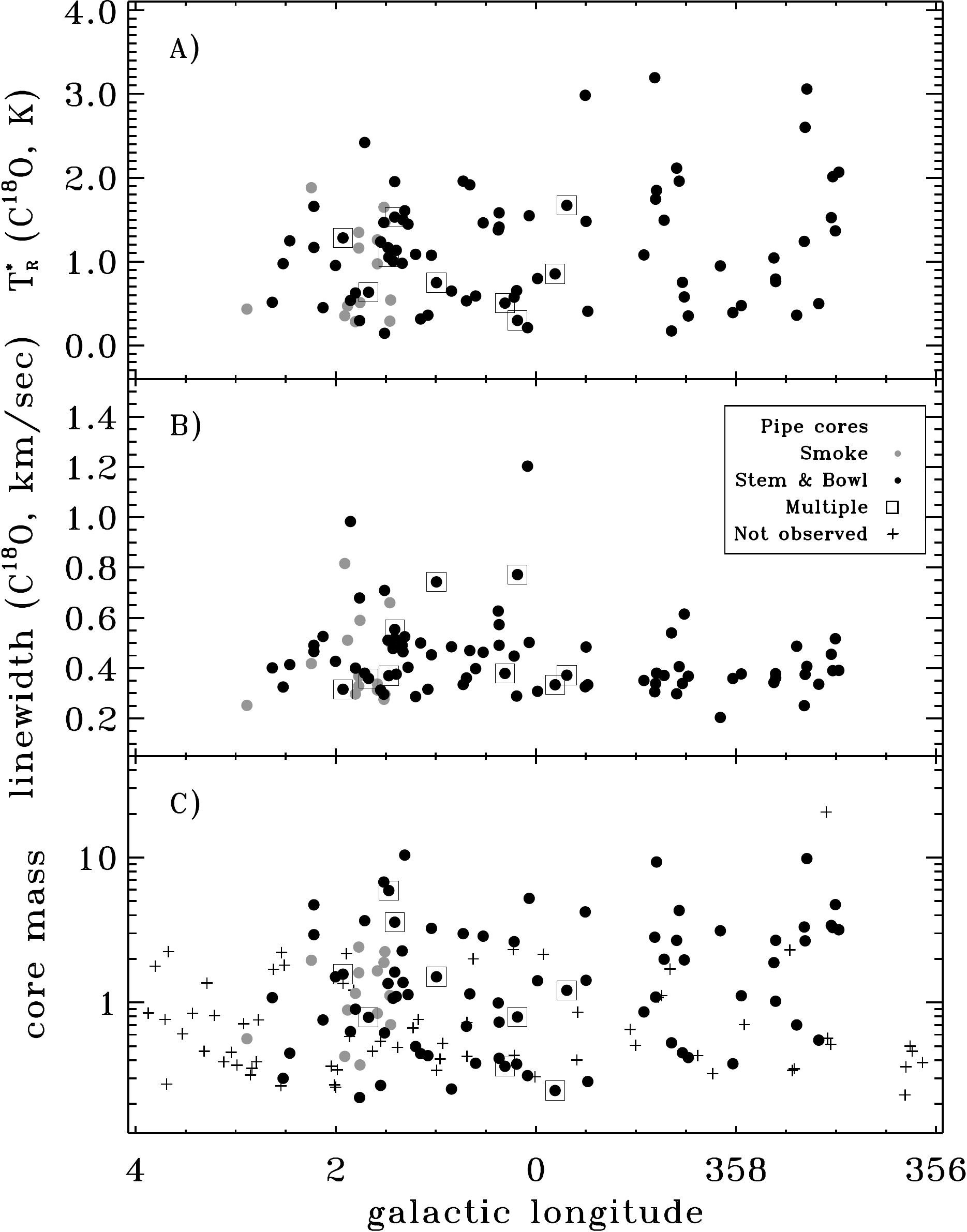}
\caption{
Homogeneity of core samples.
Core gas and dust measurements as a function of Galactic longitude.  
For multiple component cores we plot the profile fit values for the 
``a'' or brighter component.
\label{fig:data}}%
\end{figure*}%
Based upon our spatial completeness and these \vlsr\ variations we divided the
Pipe nebula into three regions. Each region was assigned a name that alludes
to the morphology of a smoking pipe~(see Fig.~\ref{fig:map1}). There is a long
$(4.2\degr; 10\;\mbox{pc})$ straight (aspect ratio, $a/b\,\sim\,3.5$) filament
tracing eastward from Barnard~59 ($l,b = 357.1\degr, 7.1\degr$), which
\citet{1927QB819.B3.......} originally called the ``Sinkhole.'' We will refer
to this very straight filament as the ``Stem'' of the Pipe. The Stem
displays extremely little variation in \vlsr\ of about 3.5~\kms. The Stem merges
with a large dark ``Bowl'' at approximately $l=0\degr$, which is where \vlsr\
increases by 1.5~\kms, though there is larger dispersion in \vlsr\ in the Bowl
compared to the Stem. Above the Bowl $(DEC\,<\,-24.75\degr)$ there are widely
dispersed smaller clouds, including the dark globules of Barnard~68, 72, 74.
We will call this region the ``Smoke;'' our \cotracer\ sample is most
incomplete in the Smoke. For reference the Smoke of the Pipe connects to an
extinction bridge (off this map) that extends $10\degr$ northward
\citep{2005PASJ...57S...1D}, connecting it to the Ophiuchus main cloud.

We used two sided KS tests to determine whether these spatial samples could be
drawn from the same parent. For these samples there a null possibility
$(10^{-8})$ that the \vlsr\ distributions for the Stem and the Bowl are drawn
from the same parent. The Smoke's $\vlsr\sim5$~\kms\ is very similar to and probably
drawn from the same parent as the Bowl; the KS probability for the Bowl and
the Smoke is 0.18. Careful inspection of Fig.~\ref{fig:map1}a, however,
reveals that bulk variations in \vlsr\ with position in the cloud are actually
more interesting in detail than we have examined here. In particular the
increased dispersion of \vlsr\ in the Bowl is due to systematic motions
(\S\ref{sec:cloud}) while most of the extinction cores that we found to
display multiple \cotracer\ components are also found in the Bowl.

We tested the extinction cores' remaining physical parameters 
$(\tant, \dvco,\mass_{core})$ for spatial homogeneity using these subsets. In
Fig.~\ref{fig:data} we plot these three physical core parameters versus
Galactic longitude. Note for simplicity the Figure only segregates the Smoke
cores by symbol; Bowl cores are those with $4\degr>l>0\degr$ 
(and $DEC>-24.75\degr$). As can be inferred visually from Fig.~\ref{fig:data}$a$, 
the 3 Pipe regions cannot be distinguished ($P_{KS}\sim0.3$) by the values 
of antenna temperature $(\tant)$.

The spatial distribution of \cotracer\ linewidths is shown in
Fig.~\ref{fig:data}$b$. While the Smoke cores could have the same parent as
either the Stem or the Bowl, the Stem and the Bowl have \dvco\ distributions
sufficiently different ($P_{KS}\sim0.002$) to warrant closer inspection. Cores
with multiple CO components were not used for these tests, although the test
results do not change when they are included. The slight differences between
the Stem and the Bowl remain even if we remove the two $\dvco>0.8$~\kms\ cores
which could also represent unresolved double-lined components. To further
explore the slight difference in linewidths between the Stem and the Bowl, we
plot \dvco\ as a function of \vlsr\ in Fig.~\ref{fig:dvvlsr}. As we have shown
that the Stem and the Bowl have distinct, distinguishable \vlsr\
distributions, it is not surprising that the slight differences in \dvco\ for
these two regions manifest themselves in a correlation with \vlsr. A linear
fits suggests \dvco\ increases by 0.05~\kms\ per unit increase in \vlsr.
As we will discuss further in \S\ref{sec:bowl} the larger linewidths within 
the Bowl are related to an interesting large scale kinematic feature of 
the Pipe nebula.

Interestingly, the distribution of extinction core masses along the length 
of the Pipe is consistent with a single parent population: $P_{KS}\sim0.3$ when 
comparing the Bowl to the Stem. Further, the mass distributions for those sample 
of cores with \cotracer\ in the three regions are also indistinguishable 
$(P_{KS}\sim0.1)$. Moreover, the mass distribution of extinction cores
that display multiple \cotracer\ components is basically uniform, and
we confirm the vast majority of these extinction cores as real, dense
cloud structures. Thus, we can conclude that the removal of such 
multi-component ``cores'' from the ensemble core mass function would not
change any of the conclusions of \citet{2007A&A...462L..17A}. 
In fact there is no quantity in our data that is
systematically contaminated by extinction cores displaying multiple \cotracer\
components.

\subsection{Correlations}
\label{sec:correlate}

\subsubsection{Core Linewidth -- Size}
\label{sec:dvsize}

\begin{figure}
\centering \includegraphics[angle=00,width=0.45\textwidth]{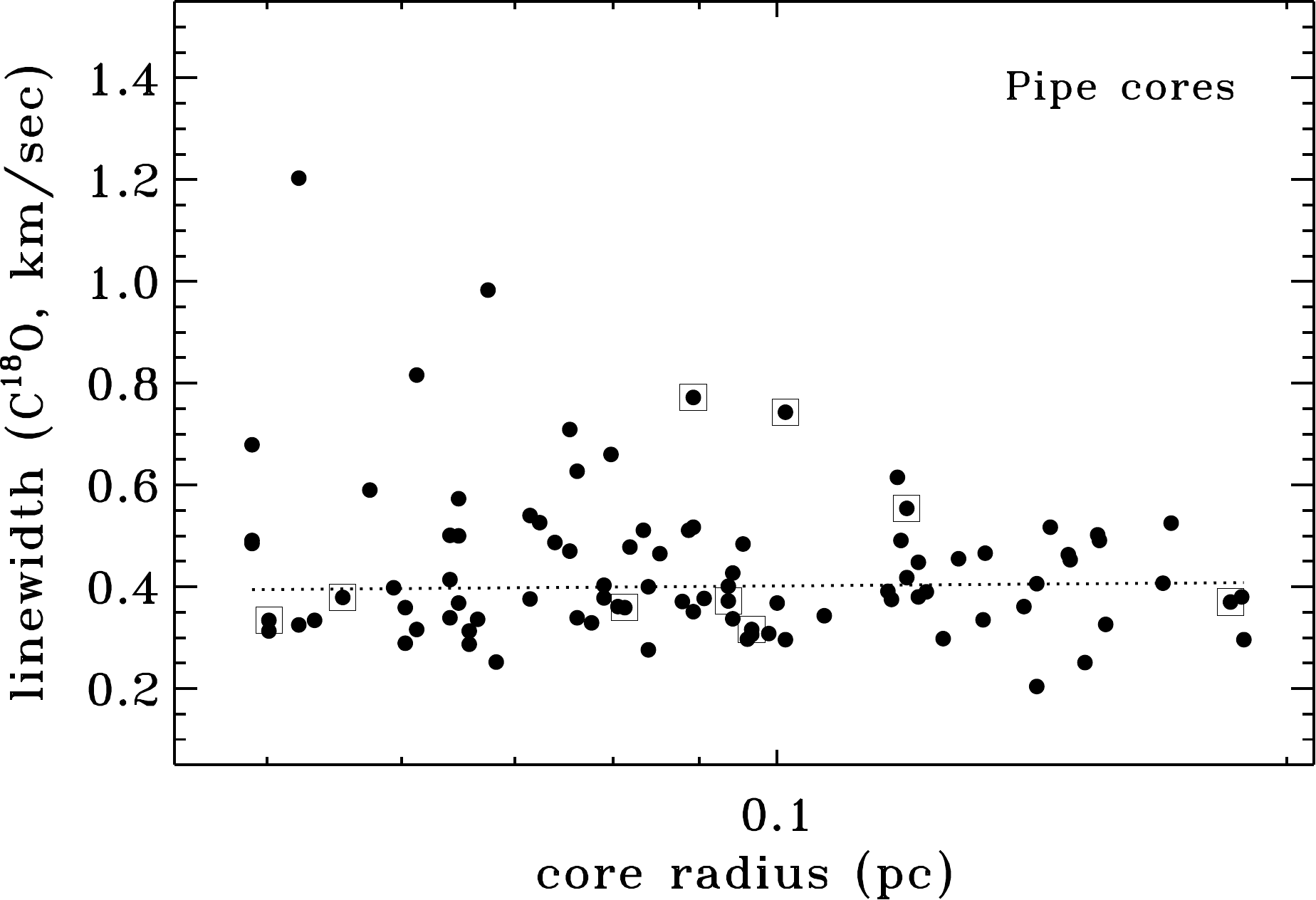}
\caption{
Core Linewidth -- Size relation for Pipe cores.
Circles surrounded by squares represent the ``a'' component
of extinction cores displaying two \cotracer\ components.
A least absolute deviation fit is shown (Table~\ref{tab:correlate}).
\label{fig:dvsize}}
\end{figure}

There is no true correlation between physical size of the Pipe extinction cores
and their \cotracer\ linewidth (Fig.~\ref{fig:dvsize}). A weak
negative correlation ($r_c\sim-0.2$) is misleading because a least absolute
deviation linear fit, which is robust against the clear outliers in this sample,
reveals nothing meaningful. All fits and correlation coefficients are given in
Table~\ref{tab:correlate}.
%
\begin{figure}
\centering \includegraphics[angle=00,width=0.45\textwidth]{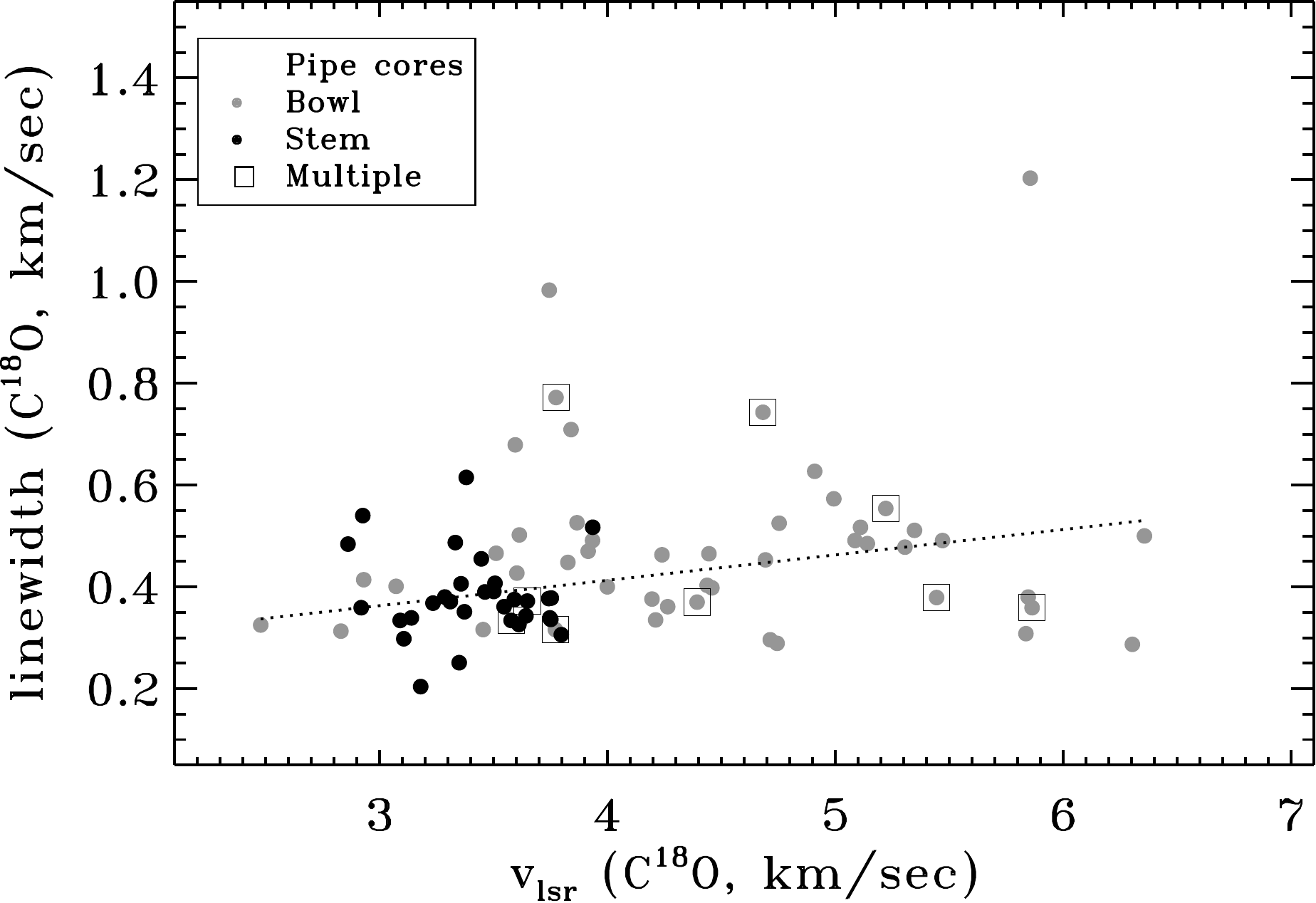}
\caption{
Correlation of \dvco\ with \vlsr.  Symbols segregated by color
correspond to cores in the Stem and Bowl of the Pipe.
Smoke cores are not included in this plot.
Circles surrounded by squares represent the ``a'' component
of extinction cores displaying two \cotracer\ components.
Note the very small dispersion in the values of both
\vlsr\ and \dvco\ in the Stem.
A least absolute deviation fit is shown (Table~\ref{tab:correlate}).
\label{fig:dvvlsr}}
\end{figure}%

%
\begin{deluxetable}{lrrr}
\tablewidth{0pt}
\tablecaption{Correlation analysis\label{tab:correlate}}
\tablehead{
\colhead{\dvco\ vs} &
\multicolumn{2}{c}{Fit\tablenotemark{a}} &
\colhead{$r_c$\tablenotemark{b}} \\
\colhead{} &
\colhead{a0} &
\colhead{a1} &
\colhead{}
}
\startdata
 $\vlsr$\       & 0.21 & 0.05 &  0.24 \\
 $\log{r}$(pc)  & 0.43 & 0.02 & -0.17 \\
 $\log{\mass_{core}}$ & 0.40 & 0.01 & -0.15 \\
 $\log{\mass_{beam}}$ & 0.47 & 0.25 &  0.08 
\enddata
\tablenotetext{a}{Result of linear least absolute 
deviation fit; a1 is the slope of the fit.}
\tablenotetext{b}{Correlation coefficient.}
\end{deluxetable}%

%

As we sample only the linewidth at the center of each extinction core it is
perhaps meaningful to ask if the \cotracer\ linewidth varies on larger angular
scales within the Pipe nebula. While variations of \vlsr\ on the largest
scales of the Pipe are discussed \S\ref{sec:cloud}, we can compare our results
to the integrated \cotracer\ linewidths data published by
\citet{1999PASJ...51..871O}. These NANTEN data were taken with a beamwidth 3
times larger than the ARO12m and their \cotracer\ data correspond only to
undersampled maps of \comap\ peaks, which miss much of the structure of the
Pipe. Indeed, we find 12 NANTEN \cotracer\ cores contain 20 extinction cores.
In 8 cases where a single NANTEN core contains only 1-2 extinction cores we
find that an additional linewidth component of order the linewidth of an
individual extinction core is needed to recover the larger scale integrated
linewidth. Determining whether this additional component comes from organized
\vlsr\ variations in the gas or from general inter-core turbulence requires
fully sampled maps. We do note that in a few cases where we have measurements
for multiple extinction cores per NANTEN core, \vlsr\ variations between
cores within the NANTEN beam are sufficient to explain these differences.

\subsubsection{Core Linewidth -- Mass}
\label{sec:linewidthmass}

%
\begin{figure*}
\begin{minipage}[c]{0.95\textwidth}
\centering \includegraphics[angle=90,width=0.90\textwidth]{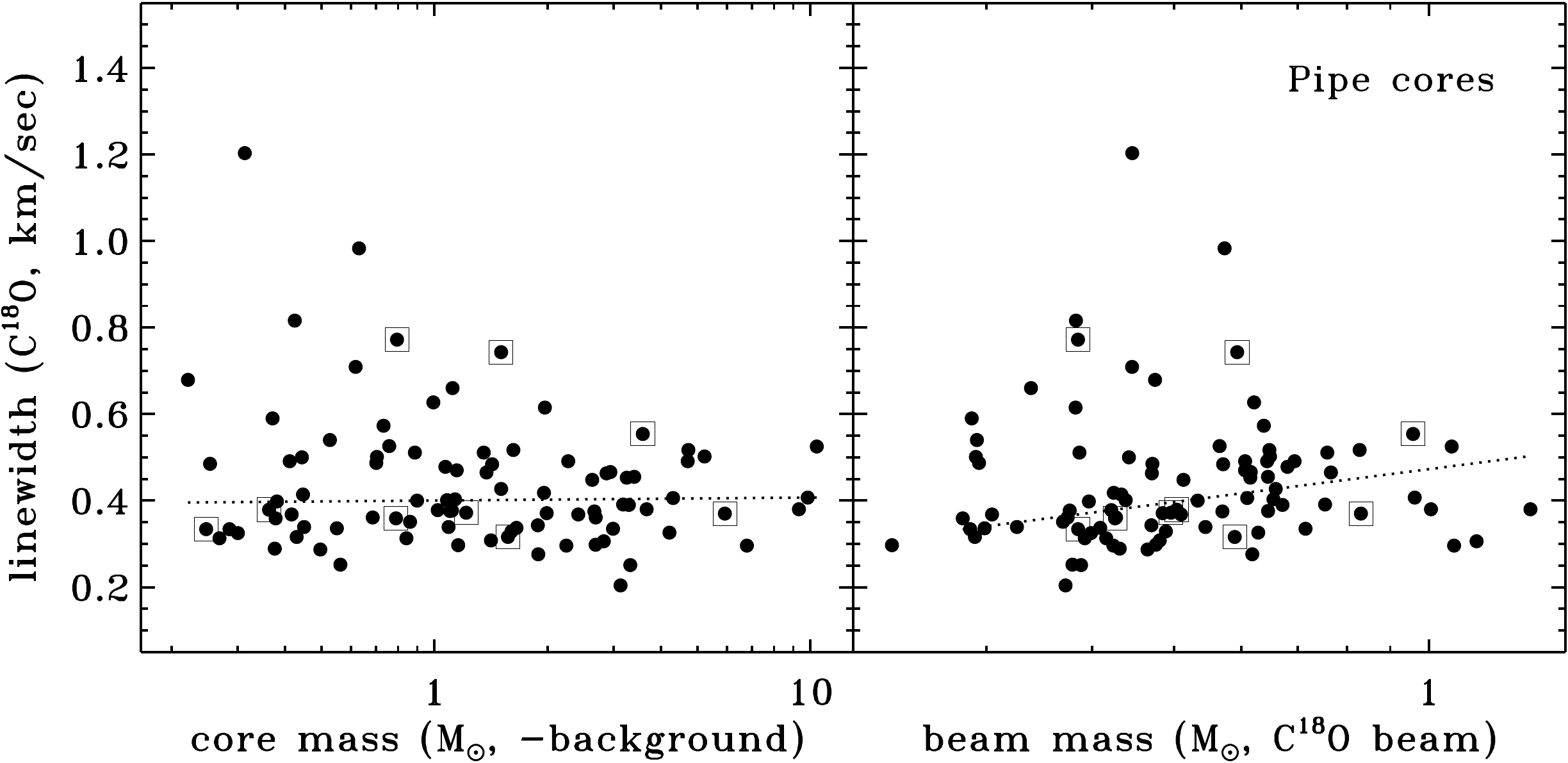}
\end{minipage}%
\caption{
Core Linewidth -- Mass relations for the Pipe.
$Left:$~Linewidth is plotted as a function of core mass, 
derived from the wavelet subtracted map of \citet{2007A&A...462L..17A}
 and tabulated in \citet[][]{CLada2007Pipe3}.
$Right:$~Linewidth is plotted as a function of the integrated column 
density seen by the radio beam and expressed in mass units. 
Beam fluxes given in Table \ref{tab:c18o}.
Circles surrounded by squares represent the ``a'' component of extinction 
cores with two \cotracer\ components.
A least absolute deviation fit is shown (Table~\ref{tab:correlate}).
\label{fig:dvmass}}
\end{figure*}
In this section we explore the dependence of the \cotracer\ linewidth on the
mass of the individual Pipe extinction cores. We use the mass of each Pipe 
extinction core as
derived from the wavelet analysis of the extinction map, which effectively
subtracts away the diffuse cloud in which the core is enveloped. As is clear in
Fig.~\ref{fig:dvmass}a there is no positive correlation between a core’s mass
and its linewidth as would be expected if the cores are in a state of virial
equilibrium. As was the case for the linewidth-size relation the scatter in
this diagram might suggest mass and linewidth are negatively correlated since
larger linewidths are more frequently observed for less massive
$(<1\solarmass)$ cores. But such an inference is due primarily to outliers as
our linear fit shows no correlation.

It may not be reasonable to assume that the linewidth at the extinction peak
should correlate with core mass. The \cotracer\ beam traces the entire column
density along the line of sight which would include any contribution from the
inter-core material whose mass is not reflected in Fig.~\ref{fig:dvmass}a because
the background was removed by the wavelet analysis. Further, for
$75\%$ of the Pipe cores the \cotracer\ beam subtends less than 20\% of the
area of the extinction core. Perhaps it is more precise to ask if the
linewidth correlates with the observed column density in the AR012m beam
$(56\arcsec)$ as determined from the 2MASS/NICER extinction map without
subtracting for background. Using {\tt wphot} in IRAF we measured the total
flux expressed in $\av$ within a Gaussian weighted beam having a $56\arcsec$\
FWHM and summed to $3\sigma\;(R=71\arcsec)$. Table \ref{tab:c18o} is
supplemented with these fluxes. In Fig.~\ref{fig:dvmass}b we convert
this flux to mass\footnote{The conversion from total column density as
measured in magnitudes of extinction $(\av)$ is $0.00757\,\solarmass/mag$. 
This constant is derived assuming a distance of 130~pc \citep{2006A&A...454..781L}, 
using a size of $30\arcsec$  per pixel in the extinction map, and employing the 
basic atomic and astrophysical constants as given by \cite{2000asqu.book.....C}. } 
and compare it to the linewidth. A linear fit suggests
that linewidth is more correlated to central beam mass than core mass.  
This could suggest that extinction cores that find themselves embedded in larger cloud 
envelopes have slightly larger linewidths than those cores which are isolated. 
Because ammonia observations that tracer higher density gas confirm the lack of linewidth
variations with the cores' size \citep{JRathborne2007Pipe2},
we instead believe that our \cotracer\ data is picking up non-negligible emission 
from inter-core gas.

\section{Discussion}
\label{sec:discuss}

\subsection{Large scale kinematics of the Pipe nebula}
\label{sec:cloud}

\begin{figure}
\centering \includegraphics[angle=90,width=0.45\textwidth]{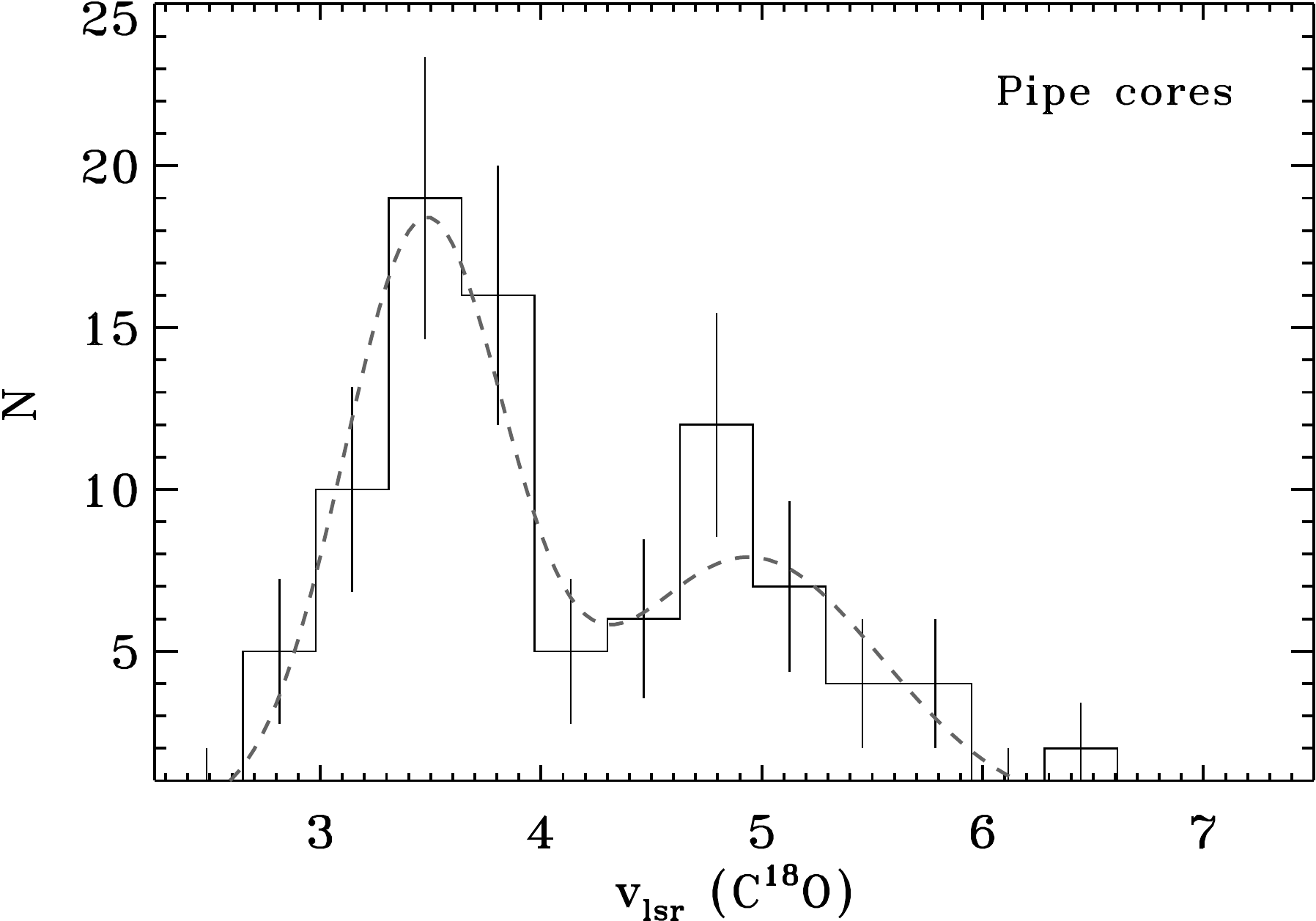}
\caption{
\vlsr\ distribution function for the Pipe \cotracer\ cores.
Two component Gaussian profile fit is shown; see text.
\label{fig:vlsr}}
\end{figure}

In this section we examine the large scale kinematics of the Pipe nebula. As
was clear in Fig.~\ref{fig:map1}, the dense extinction cores in the molecular
cloud display strong spatial \vlsr\ variations. The \vlsr\ distribution
function (Fig.~\ref{fig:vlsr}) of these cores allows us to broadly
characterize the Pipe nebula's \emph{dense} core kinematics, which appear to
consist of 2 components. Fitting these components as Gaussian distributions
yield central velocities of ${v}_{lsr,1}\,=\,3.48$ and
${v}_{lsr,2}\,=\,4.95$~\kms\ with corresponding dispersions of
$\sigma({v}_{1})\,=\,0.37$ and $\sigma({v}_{2})\,=\,0.55$~\kms.
%
%
\begin{deluxetable*}{lcccccc}
\tablewidth{0pt}
\tablecaption{Regional cloud and core extinction properties\label{tab:cloud}}
\tablehead{
\colhead{Region} &
\multicolumn{3}{c}{Cloud\tablenotemark{a}} &
\multicolumn{3}{c}{Cores\tablenotemark{b}} \\
\colhead{} &
\colhead{$N_{Pixels}$} &
\colhead{Total Flux} &
\colhead{Mass} &
\colhead{$N_{core}$} &
\colhead{$\Sigma\;\mass_{total}$} &
\colhead{$\Sigma\;\mass_{bcksb}$} \\
\colhead{} &
\colhead{} &
\colhead{$(\Sigma\av$; mag)} &
\colhead{(\mass/\solarmass)} &
\colhead{} &
\colhead{(\mass/\solarmass)} &
\colhead{(\mass/\solarmass)} 
}
\startdata
Stem                       &  68621  &  196012  & 1500 &   53 &   208 & 102 \\
Bowl                       &  83498  &  410748  & 3100 &   57 &   308 &  95 \\
Ring\tablenotemark{c}      &  24519  &  147437  & 1100 &   37 &   230 &  70 \\ 
Smoke                      & 146272  &  367895  & 2800 &   49 &   106 &  47
\enddata

\tablenotetext{a}{Cloud extinction properties derived from total
extinction map of \citet{2006A&A...454..781L} without subtraction for background.
One pixel has an angular size of $30\arcsec$. Although the regions are not circular
an effective radius for each region can be calculated as $0.01\,\cdot\,\sqrt N_{Pixels}$~pc,
assuming a distance of 130~pc \citep{2006A&A...454..781L}.
}
\tablenotetext{b}{Core masses derived by summing individual extinction 
core masses as listed in \citet{2007A&A...462L..17A} without subtraction
for the background cloud $(\Sigma\;\mass_{total})$ and 
after wavelet subtraction of the background 
\citep[$\Sigma\;\mass_{bcksb}$;][]{CLada2007Pipe3}.}
\tablenotetext{c}{Note, the Ring region is enclosed by the Bowl.}
\end{deluxetable*}%

\begin{figure*}
\centering \includegraphics[angle=00,width=0.90\textwidth]{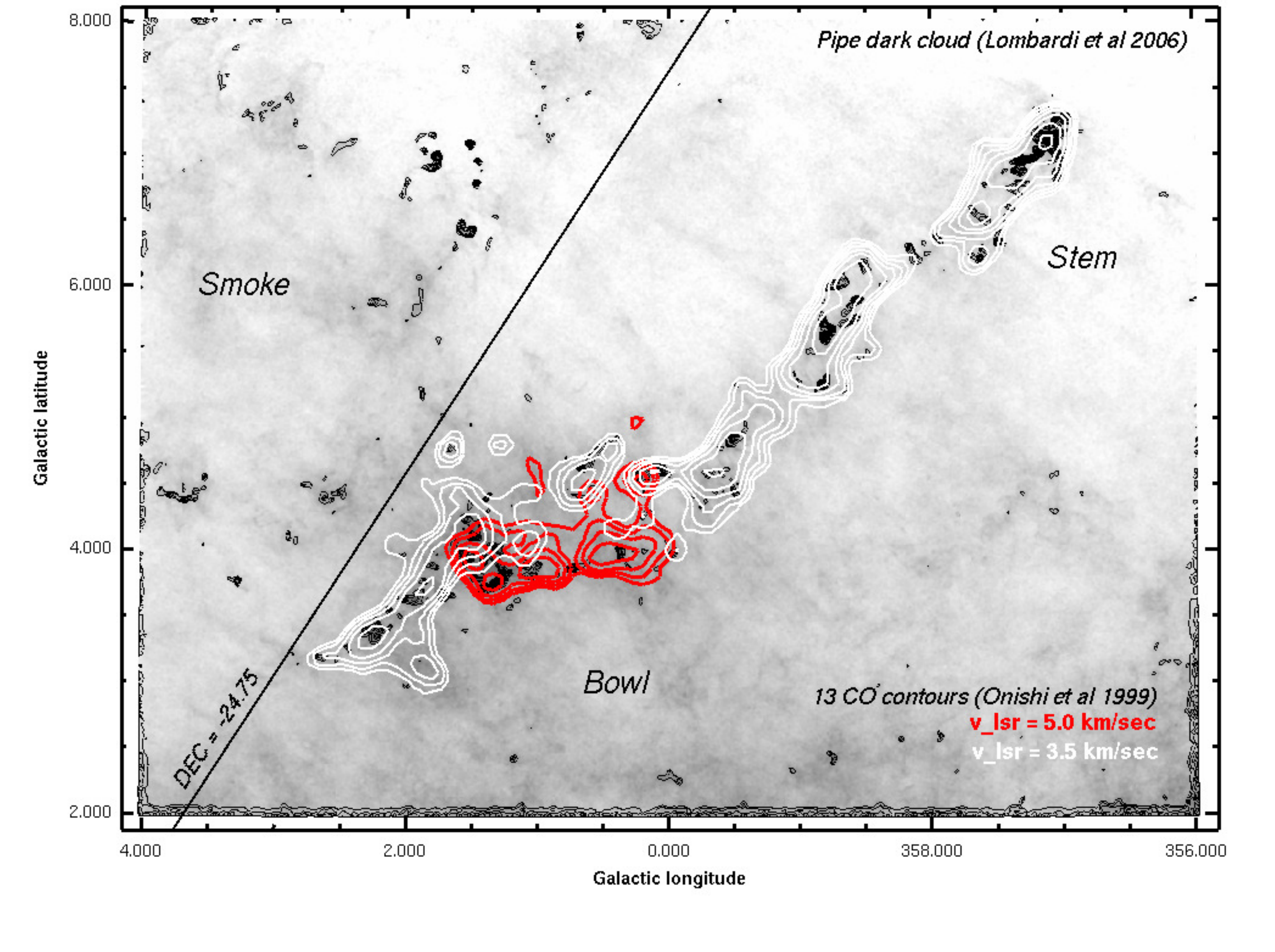}
\caption{
Smoothed \comap\ channel maps at 3.5 (white) and 5.0 (red)~\kms, 
illustrating the two primary velocity components of the Pipe nebula 
\citep[data from][]{1999PASJ...51..871O}. These components
trace the large scale kinematic feature we refer to as the 
Pipe Molecular Ring.
Black contours correspond to the wavelet subtracted map of 
\citet{2007A&A...462L..17A} overplotted on the 
complete 2MASS extinction map (inverted grayscale).
This figure was created using SAOImage DS9 \citep{2003ASPC..295..489J}.
\label{fig:map2}}
\end{figure*}

Using the \vlsr\ distribution function for the dense extinction cores as a
guide, the global kinematic structure of the Pipe cloud becomes very evident
in the \comap\ channel maps of \citet{1999PASJ...51..871O}. Overlaying two
channel maps centered at 3.5 and 5~\kms\ on to our extinction map in
Fig.~\ref{fig:map2}, we show that these two components are clearly spatially
organized: the primary component of the Pipe is a 15~pc~long filament centered
at $\vlsr\sim3.5$~\kms\ and stretching $(7\degr)$ across our entire field of view.
Surprisingly, this main component is narrow both physically and kinematically,
especially in the Stem of the Pipe. As it twists around one side of the Bowl
the main component is somewhat redshifted ($\vlsr\sim4$~\kms) relative to the 
Stem but shifts back to $3.5$~\kms\ for $l>1.5\degr$ (see also
Fig.~\ref{fig:map1}a). The molecular material of the 5~\kms\ component is
concentrated in a coherent feature within Bowl that parallels the slightly
redshifted portion of the main component. It overlaps the 3.5~\kms\ component
at $l=0\degr$ and $l=1.5\degr$, forming an apparent ``ring'' in the Bowl. We
now explore more closely the kinematics of each portion of the Pipe.
%

\subsubsection{Kinematics of the Stem}
\label{sec:stem}

As was clear in Fig.~\ref{fig:dvvlsr} the Stem of the Pipe displays a very
different kinematic signature than the Bowl; specifically, the extinction
cores in the Stem display smaller dispersions in \emph{both} linewidth and
\vlsr. The Stem consists of a single well organized very linear structure. For
Stem cores with \cotracer\ data, we can fit the \vlsr\ distribution function
with a single component, which is narrower even than the 3.5~\kms\ component
fit in Fig.~\ref{fig:vlsr}. Over its 10~parsec length the dense cores in the
Stem display a core to core standard deviation, $\sigma(\vlsr)$, of only
0.27~\kms. On the one hand the core to core motions in the Stem are slightly larger 
than the sound speed (0.2~\kms at 10K)\footnote{\citet{JRathborne2007Pipe2}
found a typical kinetic temperature of $12\pm2$ K using ammonia measurements
of a sample of these Pipe cores.}. On the other hand the gas contained within the Stem 
cores is subsonic; the median 1 dimensional \cotracer\ velocity dispersion for 
the gas within a dark extinction core in the Stem is $\sigma_{1d}=0.16$~\kms\ 
($\dvco\,=\,0.38$~\kms) with a variance of $1.3\times10^{-3}$. Summing these
gas motions, i.e., $\sigma(\vlsr)$ and $\sigma_{1d}$, in quadrature, we derive
a total 1-dimensional 
velocity dispersion for \cotracer\ in the Stem of 0.32~\kms\ or 0.56~\kms\
in 3 dimensions $(\sigma_{3d})$. Again this is not much larger than the 3-dimensional sound 
speed of molecular hydrogen (0.35~\kms). Moreover, it is significantly smaller than those 
values typical for 10~parsec structures as derived by \citet{1981MNRAS.194..809L}. 
Those structures display $\sigma_{3d}\sim2.5$~\kms.

Larson had commented that on these (10pc) physical scales large scale \vlsr\
gradients contribute significantly to $\sigma_{3d}$. By eye any gradient of \vlsr\ 
with Galactic longitude  (Fig.~\ref{fig:map1}$a$) appears quite small; 
a linear fit yields a gradient of 
$|\delta\vlsr| \sim 0.1\,\pm\,0.1\,\mbox{\kms} \mbox{deg}^{-1}$.  
The bulk motion of the gas along the Stem appears much more organized than in
those clouds described by Larson.  An alternate source for such a
missing and substantially larger non-thermal component could be the lower
density core and cloud envelopes. Our analysis of \cotracer\ on different size
scales (\S\ref{sec:dvsize}) suggests any ``additional'' non-thermal linewidth
can only be of order the typical linewidth, raising $\sigma_{3d}$
insignificantly. To examine gas at large scales and lower densities we summed
the \citeauthor{1999PASJ...51..871O} \comap\ line over the Stem; this
experiment yields a total \comap\ linewidth of $\sim1$~\kms\ or
$\sigma_{3d}\,\sim\,0.7$~\kms. We conclude that our observations of the
dark extinction cores in the Stem of the Pipe are tracing the dense
portions of a cloud in a different physical state than characterized by clouds
studied by Larson.

\subsubsection{The Bowl and a Ring}
\label{sec:bowl}

The Bowl is clearly composed of molecular gas from the two main velocity
components and it appears that these two velocity components trace a
large coherent ring-like (in projection) structure of molecular material in
the Bowl. We believe that the Pipe Molecular Ring is not a shell because there
are neither dense extinction cores nor significant \comap\ emission at any
velocity at the center of the Ring. Although the two components can be seen
clearly in the \comap\ data cube of \citet{1999PASJ...51..871O}, the existence
of multiple components in the Pipe was discussed only briefly by
\citeauthor{1999PASJ...51..871O} who noted three \cotracer\ cores at the
eastern end of the Ring that overlapped spatially but had differing values of
\vlsr.

At \vlsr=4.5~\kms\ the Pipe Molecular Ring forms a projected ellipse in \comap\
with maximum spatial dimensions of 4.7~pc by 2.9~pc and an effective radius
$(=\sqrt(area/\pi))$ of 2.1~pc. Its total mass is
$\sim1100\solarmass$~(Tab.~\ref{tab:cloud}), which was calculated from the
\citet{2006A&A...454..781L} 2MASS extinction map and excludes the flux of
the central hole having a radius $\sim1\,\mbox{pc}$ and a mass of
$150\solarmass$. \citeauthor{2007A&A...462L..17A} identified 37 dark
extinction cores in this Ring; these dense cores comprise $\sim6\%$ of the
total Ring mass. We observed 32 of them in \cotracer.
%
\begin{figure}
\centering \includegraphics[angle=00,width=0.30\textheight]{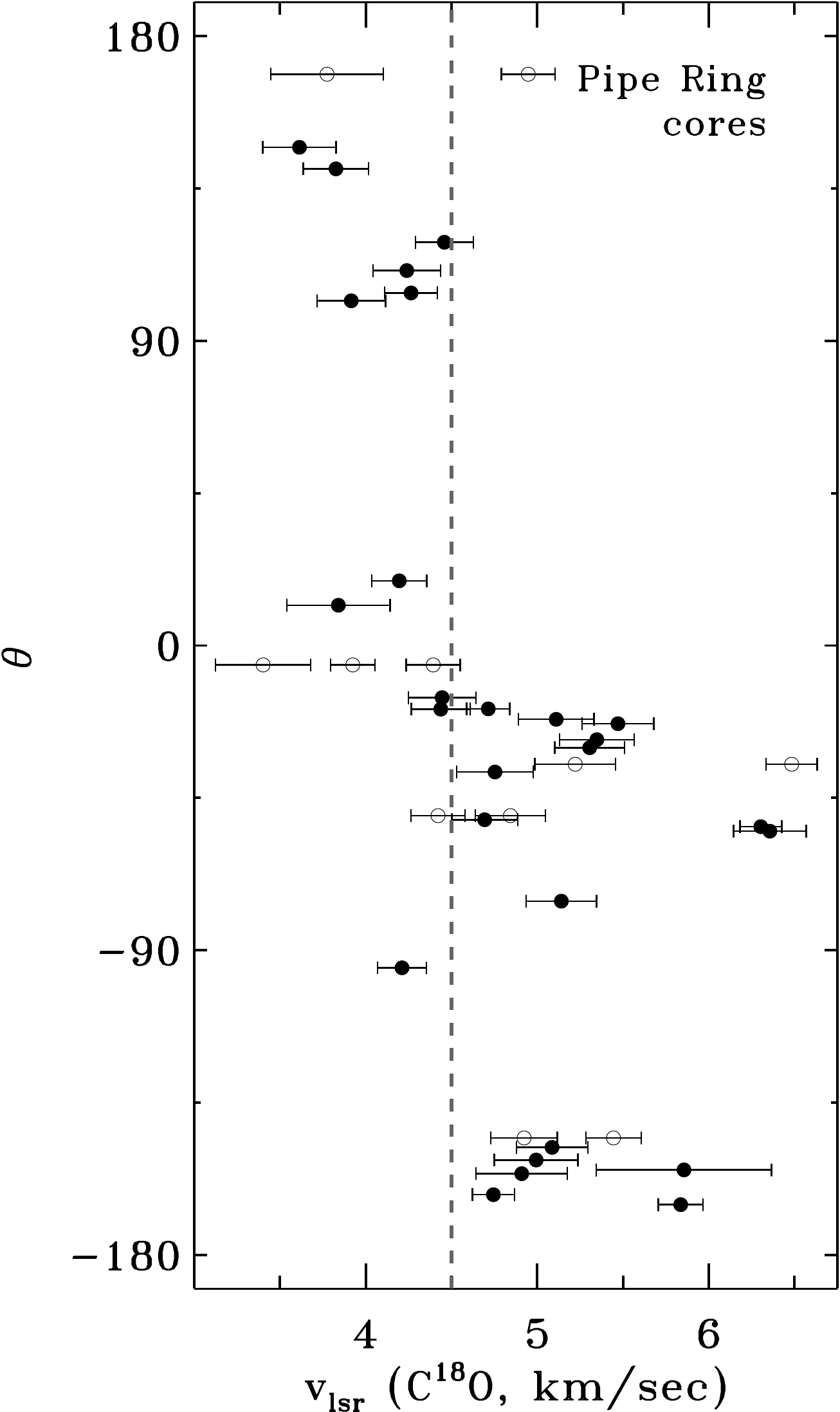}
\caption{
Variations in radial velocity within the Pipe Molecular Ring as traced by 
the \cotracer\ data of the dark extinction cores (circles).
Unfilled circles correspond to cores with multiple \cotracer\ components; 
the \vlsr\ of each component is shown.
The $1\sigma$ dispersion in \cotracer\ is overplotted for each point. 
The dashed line at $\vlsr\,=\,4.5$~\kms\ corresponds approximately
to that velocity where emission from \comap\ can be seen around 
the entire ring.
The projection to polar coordinates was performed centering on
$(l,b)_{0} = (0.755\degr, 4.172\degr)$. 
The axis of increasing Galactic longitude (East) corresponds to $0\degr$;
$90\degr$ is increasing Galactic latitude (North);
$180\degr$ corresponds to where the base of the Stem and the Ring 
meet (West, $l\sim0\degr$).
\label{fig:ring}}
\end{figure}%

The ballistics of the dense extinction cores as determined from our \cotracer\
observations permit us to further investigate the nature of this Ring. Taken 
as an ensemble the maximum radial velocity difference for the Ring cores is
2.2~\kms\ with $\sigma(\vlsr) = 0.61$~\kms. These core to core motions are 
(very) large relative to the motions of cores in the Stem (Fig.~\ref{fig:map1}) 
and are due at least in part to the systematic nature of \vlsr\ variations 
around the Ring.  These core motions are not, however, sufficient for them to
escape the gravitational field of the ring $(v_{esc}\,\sim\,2.35\,\mbox{\kms})$.
After projecting the locations of the cores into polar
coordinates\footnote{Projection centered 
$(l,b)_{0} = (0.755\degr,4.172\degr)$}, we plot the values 
of \vlsr\ versus angle, $\theta$, which
places us in the reference frame of the Ring and traces angularly the 
Ring's circumference (Fig.~\ref{fig:ring})\footnote{Note, $\theta=0\degr$ corresponds 
to increasing Galactic longitude (East); $90\degr$ is increasing Galactic latitude 
(North); $\pm180\degr$ corresponds to where the base of the Stem and the Ring 
meet (West, $l\sim0\degr$)}.
First, we can see that the dense 
cores are not behaving as a rotating ring, which would trace a parabola in 
this Figure.  There are instead strong gradients in  \vlsr\ at 
$\theta=0\degr$ and $\theta=-180\degr$. The cores in the northern
half of the Ring $(\theta>0)$ are only somewhat redshifted relative to
the main cloud $(\vlsr\,=\,4$~\kms) while their localized core to core motions 
are similar to the Stem $(\sigma(\vlsr)\sim0.27$~\kms; Table~\ref{tab:quad}).
%
\begin{deluxetable}{crccc}
\tablewidth{0pt}
\tablecaption{Ring \cotracer\ properties\label{tab:quad}}
\tablehead{
\colhead{Quadrant} &
\colhead{$N_{core}$} &
\colhead{$<\vlsr>$} &
\colhead{$\sigma(\vlsr)$} &
\colhead{$<\sigma_{1d}>$} 
}
\startdata
$180 \rightarrow 90$   &  6 & 4.05 & 0.29 & 0.19 \\
$ 90 \rightarrow 0$    &  2 & 4.02 & 0.18 & 0.23 \\
$  0 \rightarrow -90$  & 12 & 5.17 & 0.61 & 0.20 \\
$-90 \rightarrow -180$ &  7 & 5.09 & 0.54 & 0.23 \\ \hline
North                  &  8 & 4.04 & 0.27 & 0.20 \\
South                  & 19 & 5.14 & 0.59 & 0.21 
\enddata
\tablecomments{All velocity dispersions given in units of \kms.}
\end{deluxetable}%

\begin{figure}
\centering \includegraphics[angle=00,width=0.40\textheight]{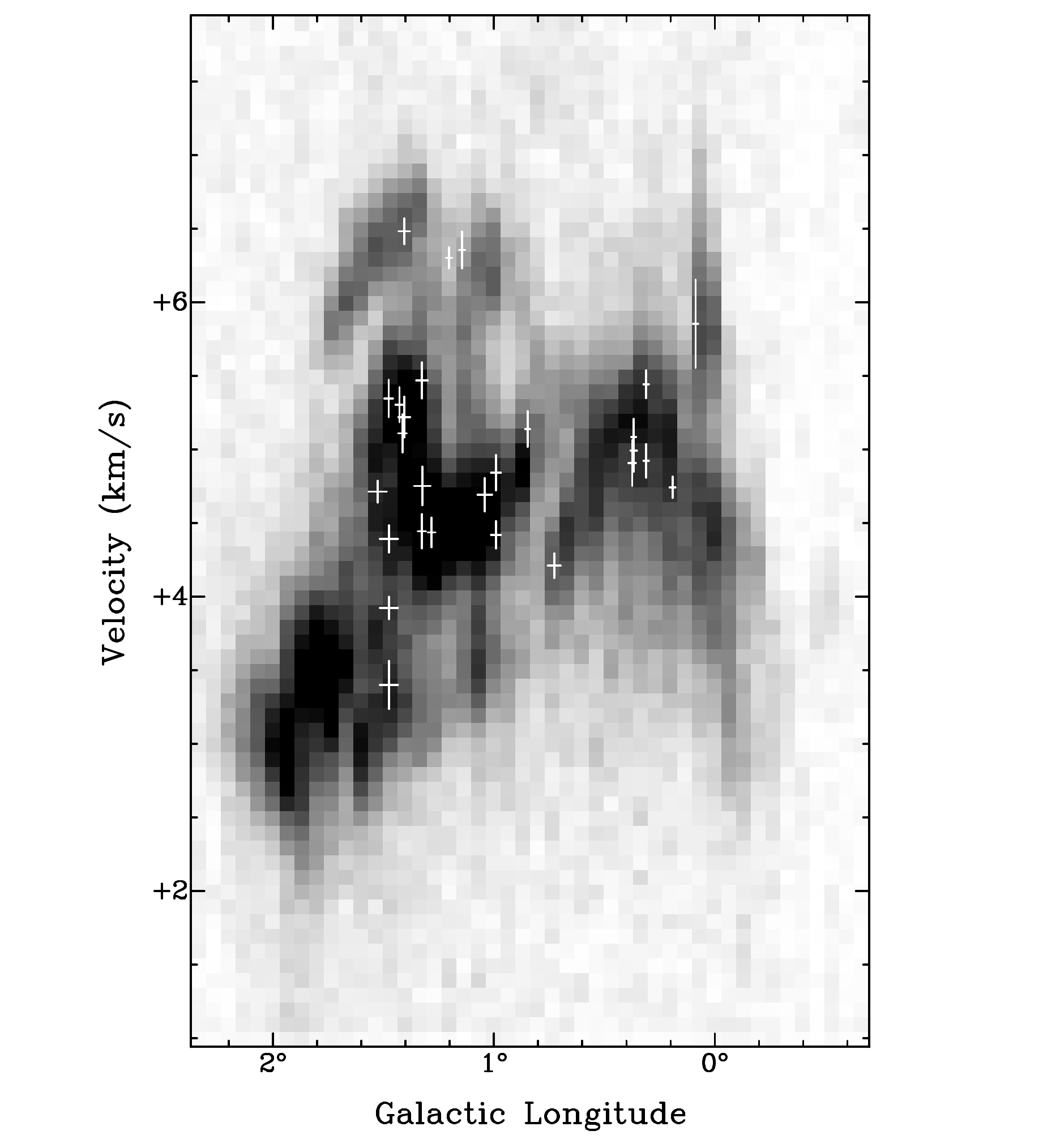}
\caption{
Variations in radial velocity along the Pipe Molecular Ring traced 
using the \comap\ data of \citet{1999PASJ...51..871O}.
This figure is the \comap\ Galactic longitude-velocity diagram $(l,v)$
averaged over a $0.6\degr$ band centered at $b=3.9\degr$.
It corresponds to the \comap\ gas enveloping those cores 
with negative values of $\theta$ in Fig.~\ref{fig:ring}.
The extinction cores in this part of the Ring are overplotted as
crosses, using their \cotracer\ \vlsr. The crosses have widths 
in $l,v$ corresponding to the extinction core's radius and
\cotracer\ linewidth.
\label{fig:south}}
\end{figure}

The extinction cores in the southern half of the ring are clearly centered on the
5~\kms\ cloud component. However, they display radial velocity variations 
($\sigma(\vlsr)\sim0.6$~\kms) that do not appear to be related to the 
overall Ring structure and are much larger than in the northern half. 
To better illustrate these motions we plot in Figure~\ref{fig:south} 
a \comap\ position-velocity  diagram aligned with this part of the 
Ring $(\theta<0\degr)$.  Note, this $l-v$ diagram is integrated 
over an $0.6\degr$ wide band centered at approximately $b=3.9\degr$. 
In \comap\ we observe a rapid but systematic transition from the 3.5~\kms\ to
the 5~\kms\ clouds at $l=1.5\degr$.  Along the extent of the 5~\kms\
component we observe large variations in the lower density gas, 
ranging from 4 to 6.5~\kms, and mirrored by the kinematics of the
denser cores.  In addition to cores at 5~\kms\ there appears to be 
gas with dense cores clustered at 6.3~\kms\ as well as at least one core 
at 4~\kms.

When we perform the same experiment for the Pipe Molecular Ring as for 
the Stem and average the \comap\ data for the Bowl, we find $\sigma_{3d}\sim1.8$~\kms,
substantially larger than the Stem.  As we have shown most of this dispersion 
comes from the systematic velocity difference between the two 
components ($\Delta\vlsr=1$~\kms) and from the larger core to core motions 
in the Ring $(\sigma(\vlsr)\sim0.6\mbox{\kms})$. Yet it also important to notice that
cores all around the Ring display linewidths elevated (0.48~\kms) 
relative to the Stem (0.38~\kms) or any other part of the Pipe 
cloud (Tab.~\ref{tab:quad}). 

It is not clear whether the Ring is a single coherent feature or
a superposition of two independent filaments, one of which 
(the $\sim5$~\kms portion) is substantially clumpier than the other.
We can say that the extinction cores in the northern portion of the Ring differ
from the main 3.5~\kms\ cloud in that they are redshifted slightly and
display slightly larger linewidths.
Even if these facts point to a physical association for the two sides
of the Ring, we cannot infer the line of sight geometry for this 
complicated region and cannot  precisely ascertain an origin for
these kinematics.
The net velocity difference $(\Delta\vlsr\,\sim\,1\;\mbox{\kms})$ 
between the two sides of the Ring may be the signature of either a
collapse or an expansion, though it does constrain the region's 
evolutionary timescale to be of order 2 Myr.
Thus, either case could correspond to a primordial kinematic feature 
from the creation of the cloud.  We prefer a primordial explanation
because the cores of the Ring appear bound to it, and we think it
unlikely that the cloud has been modified by some
localized but external event.  The kinetic energy of the ring is of 
order $10^{46}$~ergs,  which is too small to suggest the influence of 
a nearby supernovae. Such an event would likely have also similarly
impacted a much larger portion of the cloud. 
\citet{1999PASJ...51..871O} examined the 
possible influence of nearby B stars, finding them too distant to affect 
any portion of the Pipe except Barnard~59, which is $\sim5\degr$ away.  

We can further compare the Pipe Molecular Ring\footnote{We so name
this apparent feature to reflect its similarities to the Taurus Molecular 
Ring, a moniker given to HCl2 in the literature.}, to physical structures 
found in the Taurus molecular cloud. There are two features in 
Taurus that have received attention as apparent rings: \object{Heiles~Cloud~2} (HCl2) 
\citep{1984ApJ...283..129S, 1987A&A...176..299C} and \object{Barnard~18} 
\citep{1985ApJ...298..818M}. 
Both clouds have masses $(10^{3}\,\solarmass)$, radii (1-2pc) and internal 
velocity differences (1-2\kms) similar to the Pipe.
HCl2 has an obvious ring-like structure in the extinction map
of \citet{2002ApJ...580L..57P} \citep[see also][]{2004A&A...420..533T},
while Barnard~18 is primarily a weak kinematic structure without strong CO
or extinction signatures.
Moreover, the kinematics of HCl2 are much more similar to those we find in
the Pipe Molecular Ring, displaying complicated signatures of overlapping 
filaments that do not trace a rotating ring \citep{1987A&A...176..299C}. 
%

To speculate on an origin for such structures it is important to recognize a fundamental 
difference between the Pipe and Taurus, namely the present day star formation rate.
The Taurus cloud likely represents a more evolved star forming analogue to the Pipe nebula; 
indeed, both of the rings in Taurus contain young stars. In fact \citeauthor{1985ApJ...298..818M} 
suggested that Barnard~18 is an expanding ring caused by a small group of $\sim10$
stars projected near its center \citep{1982ApJ...257..620M}. As illustrated by
\citet{2004A&A...420..533T} the HCl2 ring contains star formation on
at least two sides, including a Class I protostar in \object{TMC-1} and
IRAS point sources in \object{IC~2087}. The Pipe Molecular Ring is apparently free of any
present day star formation.  \citet{1999PASJ...51..871O} 
found no IRAS point sources within the Pipe Molecular Ring that have the colors of YSOs
and our visual inspection of IRAS scans and optical DSS plates indicate to us that there
are \emph{neither} IRAS sources of any color nor bright DSS sources that could
be B stars in this Ring. Simply, there do not
appear to be any young stars in or near the Ring whose outflows would be responsible 
for its kinematics. We find that the existence of the Pipe Molecular Ring sans young 
stars supports a  hypothesis that such ring-like structures are likely primordial,
resulting from the creation of the cloud.

\subsubsection{The Smoke}
\label{sec:smoke}

Most of the Smoke extinction cores we observed with \cotracer\ are in fact members of
the 5~\kms\ component, though a full discussion of spatial variations in
\vlsr\ in the Smoke is limited by our spatial incompleteness. Excluding three
cores that are in the 3~\kms\ component, the remaining 9 \cotracer\ cores have
$\vlsr=4.8$~\kms\ with $\sigma(\vlsr)=0.2$~\kms, which is smaller even than the
Stem. The extinction cores in the Smoke have uniformly narrow linewidths, with a median
$\dvco=0.4\;(\sigma=0.14)$~\kms; their total \cotracer\
$\sigma_{3d}=0.4$~\kms\ with similar spatial dimensions as the Stem albeit
with a different projected geometry. 
As was the case for the Bowl, the \citet{2007A&A...462L..17A} cores constitute
only $2\%$ of the total column density in the Smoke (Table \ref{tab:cloud}).
Indeed, most of the mass for the Smoke is in very diffuse gas; 2/3 of the total
Smoke mass is in regions with $\av\,<\,2$~mag. Interestingly, there is no core
in the Smoke with a mass greater than $2\solarmass$.

A discussion of a few of the Smoke extinction cores is warranted. 
\object{Barnard~68}~(Pipe-101) and \object{Barnard~72}~(aka, ``the Snake,'' 
Pipe-111 and 112; \object{GF~1}~\citet{1979ApJS...41...87S}) 
can be seen at $l=2\degr; b=7\degr$ in Figs.~\ref{fig:map1} \& \ref{fig:map2}. 
These dark extinction cores have a striking, dark appearance
at optical wavelengths, which has led to their frequent appearance in
astrophotographs. These cores are separated angularly by about
$17\arcmin$ but they are separated in velocity by $1.5$~\kms\ (3.36 vs
4.96~\kms). Further observations will be valuable to determine how the 
velocities of cores in the Pipe's Smoke are spatially distributed.

\section{Conclusions}
\label{sec:conclude}

We have used the ARO12m radio telescope to characterize the
molecular gas properties of the dark extinction cores identified by
\citet{2007A&A...462L..17A} in the near-IR extinction map derived by
\citet{2006A&A...454..781L}. We obtained central pointings for 94 dark
extinction cores in \cotracer\ J=1-0 line at an angular resolution optimally
matched to the resolution of the extinction map. All the cores were detected
in \cotracer\ and less than $10\%$ of the cores display multiple velocity
components. This confirms that most of the extinction cores in the Pipe are
true dense cores and are not the superposition of unrelated filaments. We find
no correlations between the derived molecular linewidths of \cotracer\ and the
properties ($R,\mass$) of these dark extinction cores. These non-correlations
are confirmed by higher density ammonia observations
\citep{JRathborne2007Pipe2} and interpreted by \citet{CLada2007Pipe3} as
evidence that the cores of the Pipe are thermally supported and in 
pressure equilibrium with the external cloud.

While we find no correlation between basic core properties and the molecular
\cotracer\ linewidth we do find a strong correlation between a Pipe extinction
core's radial velocity and its location in the cloud. We find that there are
two well defined velocity components of the molecular gas in the Pipe. One is 
primarily located in a very narrow spatial feature that stretches 15pc across 
the region and it characterized by a small dispersion in radial velocity. 
Its internal kinematics and gas motions are inconsistent with the relations 
derived by \citet{1981MNRAS.194..809L} for large turbulent gas clouds. 
The second component forms half of an apparent ring within the Bowl of the
Pipe, having a radius of $\sim2$~pc and a mass of $\sim1000\solarmass$. 
The extinction cores in the Pipe Molecular Ring have linewidths somewhat
elevated relative those cores 
in the main cloud component.  The lower density \comap\ gas displays far larger 
motions than similar gas of the main component. Unlike similar rings in Taurus, 
there are no apparent young stars in the Pipe Molecular Ring that could be
modifying the cloud cores' kinematics.
Thus, both the Stem of the Pipe and the Pipe Molecular Ring appear to be
primordial features of this young dark cloud.
In summary we find the kinematics of the Pipe cloud to be relatively
quiescent over most of its 15~pc length having subsonic \cotracer\ extinction
cores and displaying core to core motions only somewhat larger the sound speed.
The mostly systematic nature of motion in the Pipe Molecular Ring warrants 
further detailed study through spatially complete mapping of \cotracer\
as well as molecular line transitions that trace higher densities.

\acknowledgments

We extend our appreciation to the staff of the Arizona Radio Observatory 
for supporting our observations and scheduling ample observing time.
The Kitt~Peak 12 Meter is operated by the Arizona Radio Observatory (ARO), 
Steward Observatory, University of Arizona. 
We further thank T. Onishi for access to the \comap\ cube of the Pipe. 
An anonymous referee provided comments that improved parts of this paper.
This research made use of IDL procedures developed and publicly
available from C. Markwardt. This research made use of SAOImage DS9, developed
by Smithsonian Astrophysical Observatory \citep{2003ASPC..295..489J}. This
research was supported in part by NASA Origins grant NAG10341 and NASA Spitzer
grant.

Facilities: \facility{Arizona Radio Observatory Kitt Peak 12m}


%
\clearpage
\LongTables
\begin{deluxetable*}{rcllrrrrrr}
\tablecaption{\cotracer\ data\label{tab:c18ofull}}
\tablehead{
\colhead{Pipe} &
\colhead{flag} &
\multicolumn{2}{c}{Position (J2000)} &
\colhead{$r_{sep}$} &
\colhead{$v_{lsr}$} &
\colhead{$dv$} &
\colhead{$T_{R}^{*}$} &
\colhead{rms} & 
\colhead{$\Sigma\;\av$\tablenotemark{(b)}} \\
\colhead{(ID)} &
\colhead{\tablenotemark{(a)}} &
\colhead{RA} &
\colhead{DEC} &
\colhead{$(\arcsec)$} &
\multicolumn{2}{c}{(\kms)} &
\multicolumn{2}{c}{(K)} &
\colhead{(mag.)} 
}
\startdata
    6 &     & 17:10:31.01     & -27:25:33.31    &    8.07 &   3.501 &   0.391 &   2.066 &    0.06 &    99.6  \\
    7 &     & 17:11:36.39     & -27:33:50.70    &   19.97 &   3.935 &   0.517 &   1.366 &    0.05 &    85.6  \\
    8 &     & 17:12:15.04     & -27:37:44.41    &    2.76 &   3.462 &   0.390 &   2.012 &    0.05 &    88.7  \\
   11 &     & 17:10:49.87     & -27:23:04.07    &    6.06 &   3.446 &   0.455 &   1.524 &    0.07 &    85.2  \\
   13 &     & 17:10:47.22     & -27:13:41.43    &    9.64 &   3.751 &   0.336 &   0.498 &    0.05 &    39.4  \\
   14 &     & 17:12:31.40     & -27:21:26.02    &    3.82 &   3.506 &   0.407 &   3.058 &    0.06 &   127.1  \\
   15 &     & 17:12:52.20     & -27:23:29.87    &    2.64 &   3.591 &   0.375 &   2.601 &    0.08 &    75.3  \\
   16 &     & 17:13:14.81     & -27:25:45.25    &   31.95 &   3.349 &   0.251 &   1.240 &    0.11 &    51.2  \\
   17 &     & 17:14:04.72     & -27:28:02.65    &   33.32 &   3.332 &   0.487 &   0.360 &    0.03 &    38.8  \\
   20 &     & 17:15:13.95     & -27:33:20.50    &   60.61 &   9.144 &   0.348 &   1.296 &    0.21 &    65.3  \\
   21 &     & 17:14:55.99     & -27:21:31.59    &    4.85 &   3.546 &   0.361 &   0.761 &    0.11 &    49.4  \\
   22 &     & 17:15:47.79     & -27:29:34.63    &    5.87 &   3.753 &   0.378 &   0.791 &    0.09 &    55.6  \\
   23 &     & 17:16:05.39     & -27:30:54.57    &    2.62 &   3.642 &   0.343 &   1.043 &    0.06 &    62.1  \\
   25 &     & 17:16:22.70     & -27:10:15.51    &   18.40 &   3.742 &   0.377 &   0.475 &    0.09 &    49.7  \\
   26 &     & 17:17:00.81     & -27:10:01.52    &   22.83 &   2.919 &   0.359 &   0.391 &    0.04 &    37.1  \\
   27 &     & 17:17:08.56     & -27:01:54.20    &    4.49 &   3.180 &   0.204 &   0.948 &    0.09 &    49.1  \\
   30 &     & 17:20:58.21     & -27:13:32.25    &   21.67 &   3.234 &   0.368 &   0.351 &    0.03 &    40.2  \\
   31 &     & 17:18:30.79     & -26:48:39.96    &   15.90 &   3.380 &   0.615 &   0.578 &    0.07 &    50.5  \\
   32 &     & 17:21:12.09     & -27:11:26.84    &   25.90 &   3.139 &   0.339 &   0.752 &    0.03 &    43.0  \\
   33 &     & 17:19:40.34     & -26:55:38.55    &   16.35 &   3.357 &   0.406 &   1.959 &    0.05 &    80.5  \\
   34 &     & 17:20:17.69     & -26:59:22.48    &   12.20 &   3.106 &   0.298 &   2.114 &    0.06 &    62.8  \\
   35 &     & 17:22:20.45     & -27:14:35.63    &   20.13 &   2.926 &   0.540 &   0.172 &    0.03 &    38.6  \\
   37 &     & 17:19:33.78     & -26:43:45.11    &   28.30 &   3.310 &   0.371 &   1.493 &    0.10 &    64.0  \\
   40 &     & 17:21:14.63     & -26:53:01.38    &   23.15 &   3.286 &   0.380 &   1.848 &    0.05 &   174.1  \\
   41 &     & 17:22:27.64     & -27:04:02.84    &   13.41 &   3.747 &   0.339 &   1.744 &    0.06 &    71.9  \\
   42 &     & 17:22:40.52     & -27:05:04.19    &    9.83 &   3.797 &   0.306 &   3.193 &    0.07 &   150.4  \\
   43 &     & 17:21:58.97     & -26:51:07.23    &    9.44 &   3.372 &   0.351 &   1.079 &    0.10 &    48.8  \\
   46 &     & 17:24:23.49     & -26:32:03.82    &   35.38 &   3.090 &   0.334 &   0.407 &    0.05 &    37.9  \\
   47 &     & 17:27:26.27     & -26:58:23.46    &    2.02 &   2.861 &   0.484 &   1.479 &    0.05 &    75.4  \\
   48 &     & 17:25:56.04     & -26:44:23.72    &    2.21 &   3.611 &   0.326 &   2.983 &    0.06 &    83.0  \\
   51 &   a & 17:27:22.81     & -26:44:16.14    &   10.39 &   3.648 &   0.372 &   1.670 &    0.07 &    65.5  \\
      &   b &                 &                 &         &   3.030 &   0.601 &   0.563 &         &          \\
   52 &   a & 17:28:19.47     & -26:44:02.04    &    1.77 &   3.577 &   0.334 &   0.853 &    0.04 &    50.8  \\
      &   b &                 &                 &         &   3.039 &   0.890 &   0.158 &         &          \\
   54 &     & 17:30:24.12     & -26:49:45.74    &    4.87 &   5.836 &   0.308 &   0.797 &    0.08 &    63.5  \\
   56 &     & 17:28:11.41     & -26:24:09.44    &    8.46 &   3.613 &   0.502 &   1.547 &    0.05 &    85.7  \\
   57 &     & 17:31:03.73     & -26:48:08.58    &   12.68 &   5.855 &   1.203 &   0.211 &    0.08 &    58.9  \\
   58 &   a & 17:29:41.96     & -26:29:16.61    &   11.15 &   3.774 &   0.772 &   0.298 &    0.05 &    50.8  \\
      &   b &                 &                 &         &   4.947 &   0.369 &   0.237 &         &          \\  
   59 &     & 17:30:49.24     & -26:38:34.64    &    3.42 &   4.744 &   0.289 &   0.653 &    0.10 &    56.9  \\
   61 &     & 17:28:35.63     & -26:16:47.08    &   31.72 &   3.826 &   0.448 &   0.575 &    0.07 &    67.7  \\
   63 &   2 & 17:31:34.85     & -26:36:41.42    &    5.68 &   5.181 &   0.924 &   0.458 &    0.06 &    66.5  \\
      &   a &                 &                 &         &   5.444 &   0.379 &   0.504 &         &          \\
      &   b &                 &                 &         &   4.923 &   0.457 &   0.442 &         &          \\  
   64 &     & 17:31:28.27     & -26:31:41.48    &    3.07 &   5.086 &   0.491 &   1.411 &    0.06 &    80.1  \\
   65 &     & 17:31:20.68     & -26:30:37.91    &    4.44 &   4.993 &   0.573 &   1.581 &    0.05 &    84.3  \\
   66 &     & 17:31:13.42     & -26:29:02.25    &   12.90 &   4.909 &   0.627 &   1.377 &    0.05 &    82.1  \\
   67 &     & 17:28:41.88     & -25:55:58.81    &   16.07 &   4.239 &   0.463 &   1.462 &    0.07 &    62.1  \\
   68 &     & 17:30:03.85     & -26:03:00.02    &   20.95 &   4.458 &   0.398 &   0.589 &    0.11 &    52.4  \\
   70 &     & 17:29:37.99     & -25:54:28.92    &    5.88 &   3.915 &   0.470 &   1.915 &    0.06 &    80.1  \\
   73 &     & 17:30:27.258    & -25:59:28.57    &    1.39 &   4.264 &   0.361 &   0.531 &    0.05 &    56.4  \\
   74 &     & 17:32:36.32     & -26:15:55.25    &   14.19 &   4.211 &   0.335 &   1.959 &    0.06 &    94.4  \\
   75 &     & 17:33:04.51     & -26:11:24.06    &   27.99 &   5.140 &   0.485 &   0.648 &    0.07 &    62.2  \\
   79 &   2 & 17:33:08.57     & -26:01:46.43    &   12.97 &   4.682 &   0.743 &   0.748 &    0.07 &    78.4  \\
      &   a &                 &                 &         &   4.843 &   0.481 &   0.688 &         &          \\
      &   b &                 &                 &         &   4.421 &   0.372 &   0.553 &         &          \\  
   80 &     & 17:33:32.80     & -26:01:42.21    &   13.72 &   4.693 &   0.453 &   1.075 &    0.07 &    81.2  \\
   81 &     & 17:28:36.50     & -25:15:41.55    &    7.98 &   3.454 &   0.316 &   0.360 &    0.03 &    38.4  \\
   82 &     & 17:34:31.06     & -26:02:46.66    &   32.11 &   6.356 &   0.500 &   0.315 &    0.05 &    58.4  \\
   84 &     & 17:34:51.62     & -26:01:31.82    &    5.14 &   6.303 &   0.287 &   1.086 &    0.07 &    61.4  \\
   86 &     & 17:33:27.15     & -25:43:44.12    &   26.04 &   4.437 &   0.403 &   1.446 &    0.14 &    86.5  \\
   87 &     & 17:34:30.38     & -25:49:53.16    &   84.55 &   4.753 &   0.525 &   1.606 &    0.05 &   140.5  \\
   88 &     & 17:33:47.74     & -25:43:36.70    &   39.54 &   5.470 &   0.491 &   0.979 &    0.12 &    91.6  \\
   89 &     & 17:33:28.54     & -25:40:47.98    &    6.68 &   4.445 &   0.465 &   1.497 &    0.07 &   101.2  \\
   91 &     & 17:32:13.38     & -25:25:16.92    &   20.35 &   4.195 &   0.376 &   1.134 &    0.05 &    85.2  \\
   92 &     & 17:34:04.99     & -25:39:49.30    &   36.68 &   5.110 &   0.517 &   1.953 &    0.06 &   109.4  \\
   93 &   a & 17:34:47.18     & -25:46:29.21    &   34.42 &   5.221 &   0.554 &   1.530 &    0.15 &   126.5  \\
      &   b &                 &                 &         &   6.483 &   0.351 &   0.375 &         &          \\
   94 &     & 17:34:35.50     & -25:43:16.26    &    2.97 &   5.305 &   0.478 &   1.006 &    0.06 &    89.9  \\
   95 &     & 17:22:58.55     & -23:58:03.80    &    5.53 &   4.763 &   0.501 &   0.541 &    0.03 &    38.5  \\
   96 &     & 17:23:34.99     & -24:02:22.61    &   36.19 &   4.767 &   0.660 &   0.289 &    0.03 &    44.7  \\
   97 &   2 & 17:33:30.95     & -25:30:31.25    &   20.77 &   4.128 &   1.000 &   0.870 &    0.04 &   109.8  \\
      &   a &                 &                 &         &   4.393 &   0.370 &   1.052 &         &          \\
      &   b &                 &                 &         &   3.924 &   0.305 &   0.951 &         &          \\
      &   c &                 &                 &         &   3.401 &   0.653 &   0.280 &         &          \\
   98 &     & 17:34:40.17     & -25:40:26.85    &    6.51 &   5.347 &   0.511 &   1.166 &    0.07 &   100.2  \\
   99 &     & 17:25:02.08     & -24:12:57.98    &   20.10 &   4.714 &   0.296 &   1.466 &    0.10 &    56.0  \\
  100 &     & 17:32:44.405    & -25:21:09.07    &    1.29 &   3.840 &   0.709 &   0.144 &    0.05 &    58.9  \\ 
  101 &     & 17:22:39.29     & -23:49:59.01    &   12.99 &   3.342 &   0.276 &   1.648 &    0.12 &    81.7  \\	  
  102 &     & 17:34:20.53     & -25:34:04.73    &   28.50 &   4.714 &   0.296 &   1.466 &    0.04 &   141.4  \\
  103 &     & 17:36:22.844    & -25:49:51.72    &    1.31 &   2.830 &   0.313 &   1.236 &    0.05 &    51.8  \\
  105 &     & 17:25:10.82     & -24:08:31.23    &    4.06 &   4.549 &   0.337 &   1.259 &    0.04 &    54.0  \\
  106 &     & 17:24:58.64     & -24:06:50.64    &   14.12 &   4.773 &   0.313 &   0.973 &    0.04 &    54.9  \\
  108 &   2 & 17:31:31.51     & -24:58:51.30    &   12.57 &   5.623 &   1.169 &   0.394 &    0.09 &    56.2  \\
      &   a &                 &                 &         &   5.863 &   0.359 &   0.634 &         &          \\
      &   b &                 &                 &         &   5.244 &   0.331 &   0.487 &         &          \\
  109 &     & 17:35:44.90     & -25:33:02.00    &   17.87 &   5.846 &   0.380 &   2.420 &    0.05 &   132.8  \\
  110 &     & 17:28:38.67     & -24:27:33.60    &    7.25 &   6.091 &   0.590 &   0.511 &    0.04 &    38.1  \\
  111 &     & 17:34:42.113    & -25:20:28.85    &    1.32 &   3.595 &   0.679 &   0.295 &    0.04 &    62.7  \\
  112 &     & 17:23:48.45     & -23:42:55.64    &    8.35 &   4.952 &   0.329 &   1.348 &    0.09 &    64.6  \\
  113 &     & 17:23:36.61     & -23:41:03.62    &   16.45 &   4.675 &   0.368 &   1.161 &    0.09 &    67.3  \\
  114 &     & 17:23:05.58     & -23:33:41.56    &   13.15 &   4.980 &   0.297 &   0.283 &    0.03 &    30.6  \\
  115 &     & 17:35:06.55     & -25:20:57.07    &    8.18 &   3.999 &   0.400 &   0.625 &    0.13 &    70.4  \\
  118 &     & 17:35:40.31     & -25:22:18.62    &    4.44 &   3.744 &   0.983 &   0.536 &    0.04 &    75.7  \\
  119 &     & 17:30:32.30     & -24:35:27.55    &    4.09 &   5.245 &   0.511 &   0.470 &    0.10 &    51.0  \\
  120 &     & 17:27:17.43     & -24:04:34.95    &   12.42 &   4.488 &   0.816 &   0.353 &    0.03 &    50.5  \\
  123 &   a & 17:36:22.25     & -25:23:04.26    &   26.51 &   3.771 &   0.316 &   1.282 &    0.08 &    77.9  \\
      &   b &                 &                 &         &   3.022 &   0.530 &   0.371 &         &          \\
  127 &     & 17:36:30.85     & -25:18:59.50    &    6.11 &   3.602 &   0.427 &   0.953 &    0.07 &    87.0  \\
  128 &     & 17:35:18.71     & -25:07:55.03    &   10.19 &  15.079 &   1.043 &   0.338 &    0.08 &    58.0  \\
  130 &     & 17:37:07.28     & -25:15:38.50    &   13.91 &   3.866 &   0.526 &   0.450 &    0.06 &    74.7  \\
  131 &     & 17:38:05.64     & -25:17:06.13    &    8.16 &   3.511 &   0.466 &   1.168 &    0.09 &    81.3  \\
  132 &     & 17:37:50.86     & -25:14:51.71    &   14.49 &   3.934 &   0.491 &   1.658 &    0.04 &    85.0  \\
  133 &     & 17:28:46.93     & -23:53:46.54    &   10.67 &   3.156 &   0.418 &   1.881 &    0.11 &    56.0  \\
  135 &     & 17:38:58.21     & -25:07:44.93    &    5.68 &   2.930 &   0.414 &   1.247 &    0.07 &    57.2  \\
  137 &     & 17:39:28.03     & -25:07:13.19    &    0.69 &   2.478 &   0.325 &   0.975 &    0.20 &    52.7  \\
  140 &     & 17:39:41.74     & -25:01:42.33    &   15.44 &   3.072 &   0.401 &   0.514 &    0.09 &    57.9  \\
  145 &     & 17:36:56.58     & -24:19:29.52    &   27.72 &   3.155 &   0.252 &   0.433 &    0.04 &    50.1  \\
\enddata
\tablenotetext{(a)}{Flag given to indicate \cotracer\ component for each extinction core. Components ``a'', ``b'', etc
are ordered by antenna temperature. Where given, a Flag=2, corresponds to all components fit with a single
Gaussian.}
\tablenotetext{(b)}{Total column density expressed in magnitudes of extinction (\av) measured in a Gaussian
weighted 56\arcsec\ beam on the NICER 2MASS map of \citet{2006A&A...454..781L} and centered on the 
position of the \cotracer\ observation.}
\end{deluxetable*}%


\begin{thebibliography}{26}
\expandafter\ifx\csname natexlab\endcsname\relax\def\natexlab#1{#1}\fi

\bibitem[{{Alves} {et~al.}(2007){Alves}, {Lombardi}, \&
  {Lada}}]{2007A&A...462L..17A}
{Alves}, J., {Lombardi}, M., \& {Lada}, C.~J. 2007, \aap, 462, L17

\bibitem[{{Alves} {et~al.}(2001){Alves}, {Lada}, \&
  {Lada}}]{2001Natur.409..159A}
{Alves}, J.~F., {Lada}, C.~J., \& {Lada}, E.~A. 2001, \nat, 409, 159

\bibitem[{{Barnard} {et~al.}(1927){Barnard}, {Frost}, \&
  {Calvert}}]{1927QB819.B3.......}
{Barnard}, E.~E., {Frost}, E.~B., \& {Calvert}, M.~R. 1927, A photographic
  atlas of selected regions of the Milky way ([Washington] Carnegie institution
  of Washington, 1927.)

\bibitem[{{Brooke} {et~al.}(2007){Brooke}, {Huard}, {Bourke}, {Boogert},
  {Allen}, {Blake}, {Evans}, {Harvey}, {Koerner}, {Mundy}, {Myers}, {Padgett},
  {Sargent}, {Stapelfeldt}, {van Dishoeck}, {Chapman}, {Cieza}, {Dunham},
  {Lai}, {Porras}, {Spiesman}, {Teuben}, {Young}, {Wahhaj}, \&
  {Lee}}]{2007ApJ...655..364B}
{Brooke}, T.~Y., {Huard}, T.~L., {Bourke}, T.~L., {Boogert}, A.~C.~A., {Allen},
  L.~E., {Blake}, G.~A., {Evans}, II, N.~J., {Harvey}, P.~M., {Koerner}, D.~W.,
  {Mundy}, L.~G., {Myers}, P.~C., {Padgett}, D.~L., {Sargent}, A.~I.,
  {Stapelfeldt}, K.~R., {van Dishoeck}, E.~F., {Chapman}, N., {Cieza}, L.,
  {Dunham}, M.~M., {Lai}, S.-P., {Porras}, A., {Spiesman}, W., {Teuben}, P.~J.,
  {Young}, C.~H., {Wahhaj}, Z., \& {Lee}, C.~W. 2007, \apj, 655, 364

\bibitem[{{Cambr{\'e}sy}(1999)}]{1999A&A...345..965C}
{Cambr{\'e}sy}, L. 1999, \aap, 345, 965

\bibitem[{{Cernicharo} \& {Guelin}(1987)}]{1987A&A...176..299C}
{Cernicharo}, J. \& {Guelin}, M. 1987, \aap, 176, 299

\bibitem[{{Cox}(2000)}]{2000asqu.book.....C}
{Cox}, A.~N. 2000, Allen's astrophysical quantities, ed. A.~N.~Cox (Allen's
  astrophysical quantities, 4th ed.~Publisher: New York: AIP Press; Springer,
  2000.~Editedy by Arthur N.~Cox.~ ISBN: 0387987460)

\bibitem[{{Dobashi} {et~al.}(2005){Dobashi}, {Uehara}, {Kandori}, {Sakurai},
  {Kaiden}, {Umemoto}, \& {Sato}}]{2005PASJ...57S...1D}
{Dobashi}, K., {Uehara}, H., {Kandori}, R., {Sakurai}, T., {Kaiden}, M.,
  {Umemoto}, T., \& {Sato}, F. 2005, \pasj, 57, 1

\bibitem[Kutner \& Ulich(1981)]{1981ApJ...250..341K} Kutner, M.~L., \& 
Ulich, B.~L.\ 1981, \apj, 250, 341 

\bibitem[Goodman et al.(1998)]{1998ApJ...504..223G} Goodman, A.~A., 
Barranco, J.~A., Wilner, D.~J., \& Heyer, M.~H.\ 1998, \apj, 504, 223

\bibitem[{{Joye} \& {Mandel}(2003)}]{2003ASPC..295..489J}
{Joye}, W.~A. \& {Mandel}, E. 2003, in Astronomical Society of the Pacific
  Conference Series, Vol. 295, Astronomical Data Analysis Software and Systems
  XII, ed. H.~E. {Payne}, R.~I. {Jedrzejewski}, \& R.~N. {Hook}, 489--

\bibitem[{{Lada} {et~al.}(1994){Lada}, {Lada}, {Clemens}, \&
  {Bally}}]{1994ApJ...429..694L}
{Lada}, C.~J., {Lada}, E.~A., {Clemens}, D.~P., \& {Bally}, J. 1994, \apj, 429,
  694

\bibitem[{{Lada} {et~al.}(2007){Lada}, {Muench}, {Rathborne}, {Alves}, \&
  {Lombardi}}]{CLada2007Pipe3}
{Lada}, C.~J., {Muench}, A.~A., {Rathborne}, J., {Alves}, J.~F., \& {Lombardi},
  M. 2007, \apj, submitted

\bibitem[{{Larson}(1981)}]{1981MNRAS.194..809L}
{Larson}, R.~B. 1981, \mnras, 194, 809

\bibitem[Lee \& Myers(1999)]{1999ApJS..123..233L} Lee, C.~W., \& Myers, 
P.~C.\ 1999, \apjs, 123, 233

\bibitem[{{Lee} {et~al.}(1999){Lee}, {Myers}, \&
  {Tafalla}}]{1999ApJ...526..788L}
{Lee}, C.~W., {Myers}, P.~C., \& {Tafalla}, M. 1999, \apj, 526, 788

\bibitem[{{Lee} {et~al.}(2001){Lee}, {Myers}, \&
  {Tafalla}}]{2001ApJS..136..703L}
---. 2001, \apjs, 136, 703

\bibitem[{{Lombardi} \& {Alves}(2001)}]{2001A&A...377.1023L}
{Lombardi}, M. \& {Alves}, J. 2001, \aap, 377, 1023

\bibitem[{{Lombardi} {et~al.}(2006){Lombardi}, {Alves}, \&
  {Lada}}]{2006A&A...454..781L}
{Lombardi}, M., {Alves}, J., \& {Lada}, C.~J. 2006, \aap, 454, 781

\bibitem[Motte et al.(1998)]{1998A&A...336..150M} Motte, F., Andre, P., \& 
Neri, R.\ 1998, \aap, 336, 150 

\bibitem[{{Murphy} \& {Myers}(1985)}]{1985ApJ...298..818M}
{Murphy}, D.~C. \& {Myers}, P.~C. 1985, \apj, 298, 818

\bibitem[{{Myers}(1982)}]{1982ApJ...257..620M}
{Myers}, P.~C. 1982, \apj, 257, 620

\bibitem[{{Onishi} {et~al.}(1999){Onishi}, {Kawamura}, {Abe}, {Yamaguchi},
  {Saito}, {Moriguchi}, {Mizuno}, {Ogawa}, \& {Fukui}}]{1999PASJ...51..871O}
{Onishi}, T., {Kawamura}, A., {Abe}, R., {Yamaguchi}, N., {Saito}, H.,
  {Moriguchi}, Y., {Mizuno}, A., {Ogawa}, H., \& {Fukui}, Y. 1999, \pasj, 51,
  871

\bibitem[Padoan et al.(2002)]{2002ApJ...580L..57P} Padoan, P., 
Cambr{\'e}sy, L., \& Langer, W.\ 2002, \apjl, 580, L57 

\bibitem[{{Rathborne} {et~al.}(2007){Rathborne}, {Lada}, {Muench}, {Alves}, \&
  {Lombardi}}]{JRathborne2007Pipe2}
{Rathborne}, J., {Lada}, C.~J., {Muench}, A.~A., {Alves}, J.~F., \& {Lombardi},
  M. 2007, \apj, submitted

\bibitem[{{Reipurth} {et~al.}(1996){Reipurth}, {Nyman}, \&
  {Chini}}]{1996A&A...314..258R}
{Reipurth}, B., {Nyman}, L.-A., \& {Chini}, R. 1996, \aap, 314, 258

\bibitem[{{Schloerb} \& {Snell}(1984)}]{1984ApJ...283..129S}
{Schloerb}, F.~P. \& {Snell}, R.~L. 1984, \apj, 283, 129

\bibitem[{{Schneider} \& {Elmegreen}(1979)}]{1979ApJS...41...87S}
{Schneider}, S. \& {Elmegreen}, B.~G. 1979, \apjs, 41, 87

\bibitem[{{Tachihara} {et~al.}(2000){Tachihara}, {Mizuno}, \&
  {Fukui}}]{2000ApJ...528..817T}
{Tachihara}, K., {Mizuno}, A., \& {Fukui}, Y. 2000, \apj, 528, 817

\bibitem[{{T{\'o}th} {et~al.}(2004){T{\'o}th}, {Haas}, {Lemke}, {Mattila}, \&
  {Onishi}}]{2004A&A...420..533T}
{T{\'o}th}, L.~V., {Haas}, M., {Lemke}, D., {Mattila}, K., \& {Onishi}, T.
  2004, \aap, 420, 533

\bibitem[Williams et al.(1994)]{1994ApJ...428..693W} Williams, J.~P., de 
Geus, E.~J., \& Blitz, L.\ 1994, \apj, 428, 693

\bibitem[{{Wolf}(1923)}]{1923AN....219..109W}
{Wolf}, M. 1923, Astronomische Nachrichten, 219, 109

\end{thebibliography}
\end{document}